\documentclass[twoside,11pt]{article}

\usepackage{amssymb, amsmath}
\usepackage{graphicx, subfigure, wrapfig}
\usepackage{amsfonts}
\usepackage[english]{babel}

\def\0{{\bf 0}}

\def\p{{\mathbf{p}}}
\def\q{{\mathbf{q}}}

\def\v{{\bf v}}

\def\x{{\bf x}}

\newtheorem{theorem}{Theorem}[section]
\newtheorem{remark}[theorem]{Remark}
\newtheorem{proposition}[theorem]{Proposition}
\newtheorem{corollary}[theorem]{Corollary}
\newtheorem{definition}[theorem]{Definition}
\newtheorem{conjecture}[theorem]{Conjecture}

\begin{document}


\title{On  the $n$-body problem   on surfaces of revolution}

\author{Cristina Stoica\\
Department of Mathematics\\
Wilfrid Laurier University, Canada\\
Email: cstoica@wlu.ca}

\maketitle

\noindent
Abstract: We explore the $n$-body problem, $n\geq 3,$ on a surface of revolution with a general  interaction depending on the pairwise  geodesic distance. Using the geometric methods of classical mechanics we determine a large set of properties. In particular, we show that Saari's conjecture fails on surfaces of revolution admitting a geodesic circle. We define homographic motions and,  using  the discrete symmetries, prove that when the masses are equal, they form an invariant manifold. On this manifold the  dynamics are reducible to a one-degree of freedom system. We also find that for  attractive  interactions, regular $n$-gon shaped relative equilibria  with trajectories located on geodesic circles typically experience a pitchfork bifurcation. Some applications are included.

\vspace{1cm}



\tableofcontents

\section{Introduction}

There are many generalizations of the classical $n$-body problem. 
%
%
 For instance, one may modify or  generalize the  interaction potential (\cite{Boc96}), or enlarge   the configuration space to a higher dimensional  Euclidean space  (\cite{Alb98}), or  endow the configuration space with a  non-Euclidean   structure (\cite{Mont15, Sch06b}). In particular,  in the last decade a good body of work was dedicated  to the study of the \textit{curved $n$-body problem}, that is the  generalization of the classical $n$-body problem  to surfaces of constant curvature (\cite{{Bor04}, Diacu2011, Diacu2012, Diacu16, Fra16, Gar16, Mart13, Per12, Tib13,  Sch06, Vo05}). Also recently, a unified formulation of $n$-body and $n$-vortex dynamics on Riemann surfaces was advanced, bringing together two related but quite distinct  systems (\cite{Boa16}).

In this paper we report  findings on the generalization of the $n$-body problem, $n\geq 3$, to  a system  confined  to a surface of revolution. The mass points are interacting pairwise via some potential depending on the  shortest geodesic path. The potential  
 remains unspecified until the final section of the paper when we present some examples.
%

Using  the geometric methods of classical mechanics, we find  a large set of   properties. We retrieve the integrals of motion, find which   geodesics  are configurations of  invariant manifolds,   and give a negative answer to the generalization of Saari's conjecture on surfaces of revolution admitting at least one geodesic circle (e.g. a 2-sphere). 
Using discrete symmetries, we also  prove  that  when all masses are equal, homographic motions (defined below) form an invariant manifold, and  recognize a generalization of the Clairaut relation on surfaces of revolution. One of the most interesting properties  is that  for motions with (unspecified)  attractive interaction, regular $n$-gon shaped  relative equilibria  with trajectories  located on a geodesic  circles normal to the rotation axis typically  undergo a  pitchfork bifurcation.  To our knowledge, this property   was not  remarked upon  even in particular cases such as the  gravitational equal mass  $3$-body problem on a 2-sphere (see, for instance, \cite{Diacu16, Mart13}).

   The  generalized $n$-body problem on a surface of revolution   is a mechanical system with $2n$ degrees of freedom  symmetric with respect to rotations about the axis of revolution.
  We model the dynamics  in   Lagrangian  and  Hamiltonian formulations. Considering the axis of revolution  of the configuration surface as vertical, we remind the reader that the geodesics of a surface of revolution are either meridians, parallel circles at ``heights" at which the generatrix has a critical point,  or Clairaut curves. (see Section \ref{sect: surf_rev} or \cite{ONeil}).  We prove that motions with initial positions on, and velocities tangent  to,  parallel geodesic circles or meridians are invariant manifolds, a result natural from a physics standpoint.  While one might expect that motions aligned to  Clairaut curves  also form invariant manifolds, this is true only locally.
  We retrieve  the  conservation of  energy and  angular momentum, the latter due to the invariance of the dynamics to rotations about the axis of revolution. We also calculate the  moment of inertia associated to the rotational symmetry of the surface, and give some criteria for  the existence of relative equilibria (RE). We  observe that the dynamics on a cylinder exhibits of an extra integral, namely the  linear momentum along the vertical; moreover, an associated ``vertical centre of mass" integral is also present.

Recall    the so-called \textit{Saari's conjecture}: proposed in 1970 by Don Saari in the context of the planar $n$-body problem, it claims that solutions with constant moment of inertia are RE. 
This conjecture has been proven for the planar 3-body problem with equal masses in \cite{McC04} and for the general planar $3$-body problem in \cite{Moe05}. 
Several researchers have worked on various restrictions or generalizations of this conjecture (\cite{DPS2005, Roberts2006, Her05, LaSt07, ScSt2006}). In recent years,   it was tackled by  Diacu et al. (\cite{Diacu2012}) in the context of  the $3$-body problem on surfaces of constant curvature. In this paper we consider the natural generalization of the Saari conjecture to  $n$-body problems on surfaces of revolution. 
%
The definition  of moment of inertia is generalized in this context using
%
 using  the more general geometric mechanics notion of locked inertia tensor, which in this case is a scalar quantity.
 The history of the moment of inertia  as understood in celestial versus classical mechanics may be found in \cite{Diacu2017}.
We show  that the Saari's conjecture  on surfaces of revolution which admit at least one geodesic parallel circle is not true. The result is independent of the potential or the curvature of the surface and, in particular, it settles the   generalization of Saari's Conjecture  on a 2-sphere.

By definition, in the classical $n$-body problem,  homographic solutions are those for which \textit{the configuration formed by the bodies at a given time moves in the inertial barycentric coordinates system in such a way as to remain similar to itself when $t$ varies}  - see \cite{Win41}.  
Without the conservation of linear momentum and consequently, in the absence of a barycentric system, homographic solutions are defined as those with trajectories forming a self-similar shape in the ambient $\mathbb{R}^3$ space (see \cite{Diacu2012T}). 
Here we  define homographic motions as the set of homographic solutions  which, besides maintaing a self-similar shape in  $\mathbb{R}^3$, have the points located on a plane orthogonal to the axis of revolution at all times.
Homographic motions   include as special cases homothetic motions (the bodies move without rotations along the surface meridians), and  RE.

Recall that for an ODE system with discrete symmetries, an effective method to obtain invariant manifolds of solutions consists in restricting the dynamics to the associated fixed point spaces (see, for instance, \cite{Gol88}).
  This method, known as \textit{discrete reduction}, was specialized to the  symplectic and  the cotangent bundle categories (\cite{GS84, Ma92, Montaldi98}), with the outcome that   the fixed point spaces are    symplectic and cotangent bundle systems, respectively. Also, by Palais' Principle of Symmetric Criticality (\cite{Palais68}), any RE in a fixed point space is a RE in the full phase space. Moreover, one can demonstrate the  lack of stability of certain RE, by proving that an RE  is unstable on  a fixed point space  (and thus is unstable in the full phase-space).

When the masses are equal,  we apply  Discrete Reduction  to show 
the presence of  homographic motions. In our context, these form a symplectic  invariant manifold with solutions for which   the configuration of the bodies maintains  a regular  polygonal shape at all times. We recognize the  homographic dynamics as given by  a two degrees of freedom system of the form ``kinetic + potential" with rotational symmetry.  We then show that equal mass homographic solutions obey a generalized Clairaut relation, which is essentially a reformulation of the angular momentum conservation.   
Further,  using the conservation of angular momentum, we reduce the dynamics to a   one degree of freedom, and thus integrable, system parametrized by the energy and the angular momentum.  
As known, for such systems  a sketch  of the  amended potential  is sufficient to provide   a complete qualitative picture of the dynamics (\cite{Arnold78, Sm70}).
%
Consequently,    if the  generatrix of the surface of revolution and the binary potential  are specified,  then one is able to completely  describe the dynamics on the equal mass homographic  invariant manifold, both  quantitatively (i.e. the solutions up to some quadrature) and qualitatively (i.e. the  topological portrait of the phase space and all  orbit types).   

We remark that the same problem was recently considered in a purely mathematical context by Fomenko et al. \cite{Fom15}:  
 the geodesic flow on a surface of revolution 
augmented by a rotationally symmetric potential. This flow  can be recognized  as the  homographic flow for the equal mass $n$-body problem (modulo a straight-forward scaling) on surface of revolution. 
In this interesting study,  the authors provide  
the  topological picture of the phase space  in the terms of the so-called Fomenko-Zieschang invariants.

We next study \textit{Lagrangian homographic RE}, that is  RE  which are solutions on the  equal mass homographic invariant manifold with their configuration on a plane orthogonal to the symmetry axis. These RE  maintain  a regular $n$-gon configuration  and rotate with (an appropriate) constant angular velocity on a parallel circle. After some existence criteria,  we state and prove one of the main results in this paper (Proposition \ref{pitchfork} and Remark \ref{typical}):  if the  binary interaction is attractive then a Lagrangian homographic RE with  its trajectory on a geodesic  circle typically undergoes a  pitchfork bifurcation. The  bifurcation parameter is the  angular momentum.

 We choose to not  specify any potentials  until the last section, where  we demonstrate  our   theoretical findings on  some examples. 
  First  we choose  an attractive  interaction, which we call \textit{quasi-harmonic},  given by a  pairwise potential  of the form  $G(x)=x^2/2$, where $x$ is the (geodesic) distance between two unit mass points. This potential  has the   exceptional feature  that the sum of the potential terms can be collapsed to a single term. (A $n$-body potential is a sum of $n(n-1)/2$ terms; in general, this sum does not collapse.) 
  Thus,   in our examples we obtain  the bifurcation momenta and their location  as  expressions  (depending on the generatrix) for general $n$. Using this potential, we describe the homographic dynamics for   motions on the unit sphere $\mathbb{S}^2$, a symmetric peanut-like surface and a one-sheet hyperboloid  $\mathbb{H}^2_{\text{one}}$ with a geodesic circle of unit radius .   

Next, we consider   homographic  motions on $\mathbb{S}^2$ and  $\mathbb{H}^2_{\text{one}}$ with 3-d gravitational potentials. Recall that the later  were found as    solutions of the Laplace equation on 3-d surfaces of constant curvature and then restricted to $\mathbb{S}^2$ and  $\mathbb{H}^2_{\text{one}}$ (for more on gravitational potentials and historical notes, see for instance, \cite{Diacu2011, Vo05}). These potentials  are 
\begin{equation}
G(x)=- m_i m_j \cot x \,\,\,  \text{on}  \,\,\,  \mathbb{S}^2 
\quad \quad \text{and}\quad \quad 
G(x)=- m_i m_j \coth x   \,\,\,  \text{on}  \,\,\,  \mathbb{H}^2_{\text{one}}
\label{grav_pot}
\end{equation}
where $x$ is the (geodesic) distance between two points of mass $m_i$ and $m_j$. Note that for the potential on $\mathbb{S}^2$, configurations with diametrically opposite points are ill-defined; in particular,  for motions on $\mathbb{S}^2$, we thus consider Lagrangian homographic RE for $n$ odd only.  
Using Proposition \ref{pitchfork} we find the condition which guarantees that a Lagrangian homographic RE  with its trajectory on the Equator  will  experience a subcritical pitchfork bifurcation. (To simplify exposition, we also call Equator the geodesic circle of $\mathbb{H}^2_{\text{one}}$.) We verify this condition for $n=3$ and conjecture that it will be fulfilled for any $n\geq 5$ odd.
%

A physically relevant harmonic or gravitational potential on a general surface of revolution varies with the curvature. The gravitational case was discussed by  Santoprete  \cite{Santoprete07}, whereas the harmonic case awaits investigation.
    (As a remark, it would be interesting to find the proper  harmonic potential on surfaces of revolution using a  symmetry Cayley-Klein-type approach  such as in the paper of Cari$\tilde{\text{n}}$ena et al \cite{CR05}). 
The work we present here focuses on finding generic properties for the $n$-body problem of revolution, leaving for the future   more involved studies for specific potentials.

  \bigskip
   The paper is organized as follows: in Section 2 we set up the problem in Lagrangian and Hamiltonian formulations. We discuss geodesics as invariant manifolds, deduce  the conservation laws and  the RE existence conditions, and observe on the existence of an additional  conservation law for motions on a cylinder.
In Section 3 we  prove that the generalization of Saari's conjecture  fails on surfaces of revolution with at least one geodesic parallel circle. In Section 4 we  define  homographic motions,  show that they form an  invariant manifold, and state and prove a generalization of the Clairaut relation. Further, since homographic motions are given by a two degree of freedom Hamiltonian system of the form ``kinetic +potential" with rotational symmetry, we reduce the dynamics to a one degree of freedom system and remind the reader that a full qualitative picture of the dynamics is provided by the analysis of the amended  potential. In Section 5 we retrieve some RE existence criteria, and state and prove  Proposition  \ref{pitchfork} on the bifurcations of RE with trajectories on a geodesic circle. In Section 6 we discuss apply our findings on some examples for the quasi-harmonic and   gravitational interactions.



\section{Motion on a surface of revolution}

\subsection{Surfaces of revolution}
\label{sect: surf_rev}

Consider  a  surface of revolution $\mathbb{M}$ generated by  rotations  about the vertical $Oz$ axis of a profile smooth curve $f=f(\cdot)$ defined on some open $(-a,b)$ interval with $a$ and $b$ finite or  $ \infty$.  We chose to parametrize $\mathbb{M}$ by
\begin{align}
(z, \varphi) \to 
\x(z, \varphi):=\left(f(z)\cos \varphi, f(z)\sin \varphi, z  \right)
\label{param_M}
\end{align}
with $z\in (-a,b)\,,\, \varphi \in {\mathbb S}^1$, where $z \to f(z)$ is the given smooth profile curve  with $f(z)>0$.  
Recall that the geodesics of a surface of revolution  are:

\smallskip
\noindent
- \textit{geodesic circles}, that is parallel circles $z=z_{c}=$constant,  where $z_{c}$ is a  critical point  of the generatrix $f(z)$, and 

\smallskip
\noindent
- \textit{Clairaut curves}, that is curves $\varphi=\varphi(z)$  that are solutions of ODEs 
 \begin{equation}
\frac{d\varphi}{dz}= \frac{c}{f(z)}\frac{\sqrt{f'^2(z) +1}}{\sqrt{f^2(z)-c^2}}
\label{geodesic}
\end{equation}
where 
\begin{equation}
f^2(z) \dot \varphi(z)=f^2(z_0) \dot \varphi(z_0)=const.=c\,.
\label{ang_mom_geo}
\end{equation}
%
%
The latter category includes meridians $\varphi=$ constant obtained for $c=0.$ The relation \eqref{ang_mom_geo} is the expression of the  angular momentum for the dynamics conservation associated to the geodesic equations. 
For future reference, we the same relation written as 
\begin{equation}
f(z) \cos \theta = c
\label{Clairaut_geo}
\end{equation}
where $\theta\in (0\,,\pi/2)$ is the angle between the geodesic and a parallel circle  is known as  the \textit{Clairaut relation} (see \cite{do_Marmo76}, pp. 257). 

\bigskip
On $\mathbb{M}$ we define the distance function
$
d: \mathbb{M} \times \mathbb{M} \longrightarrow \mathbb{R}$\,,
$d\,(\q_1, \q_2) :=$ \textit{the} \textit{shortest} \textit{distance} \textit{from} $\q_1$ \textit{to}  $\q_2$ \textit{along} \textit{a} \textit{geodesic} \textit{arc}\,. 
We observe that  in the parametrization above 
\begin{align}
d: (-a, b) \times (-a, b) \times \mathcal{S}^1 \times \mathcal{S}^1 &\to [0, \infty) \nonumber\\
\,\,\,\,\,\left( \,z_1\,, \,z_2\,, \,\varphi_1\,, \varphi_2 \,\right) &\to d\,\,(z_1, z_2, \,\varphi_1\,, \varphi_2)\,,
\label{dist_function}
\end{align}
and that  $d$ is  rotationally invariant, that is
\begin{equation}
d\,\,\left(z_1, z_2, \varphi_1\,, \varphi_2\right) =d\,\,\left(z_2, z_1, (\varphi_2-\varphi_1) \right)\,.
\end{equation}
%
%
%
%
%
%
%
%
%



\subsection{Lagrangian formulation and geodesics as  invariant manifolds}

Consider $n$ mass points  $P_i$  of mass $m_i$  on $\mathbb{M}$ that are  interacting mutually via a potential depending on the (shortest) geodesic distance between the points.
Denote 
the coordinates of  $P_i$ by $\q_i=(z_i, \varphi_i)$, $i=1,2\ldots, n$, and let $\q:=(\q_1, \q_2, \ldots, \q_n).$ The configurations space   is
${\mathbb{Q}}:= {\mathbb{M}}^n \setminus \left\{ \right.$collisions and configurations where the vector field is undefined
 (see Remark \ref{config_space} below)$\left. \right\}$.
The dynamics is given by the  Lagrangian $L:T{\mathbb{Q}} \to \mathbb{R}$
\begin{equation}
L(\q, \dot \q) = T_\q(\dot \q) -V(\q)
\label{Lagrangian}
\end{equation}
where $T$ is a mass-weighted metric on ${\mathbb{M}}^n$ defined by
\begin{equation}
T_\q(\dot \q) =\sum\limits_{i=1}^n m_i \left< \dot \q_i\,, \dot \q_i \right>:= 
\sum\limits_{i=1}^N \frac{1}{2} m_i 
\left[\dot z_i\,\, \dot \varphi_i \right]
\left[
\begin{array}{cc}
1 + f'^2(z_i) & 0 \\
0 & f^2(z_i)
\end{array}
\right]
 \left[
\begin{array}{c}
\dot z_i \\
\dot \varphi_i
\end{array}
\right]
\label{metric}
\end{equation}
and 
$V(\q)$ is the interaction potential. We assume that any two points, say $P_i$ and $P_j$ of coordinates $\q_i$ and $\q_j,$ respectively, interact via a law depending on  $d(\q_i, \q_j)$. 
Further,  
we assume that the potential $V$ is either independent of the masses, and so
\begin{equation}
V(\q):=\sum \limits_{1\leq i< j\leq  n} G(d (\q_i, \q_j) ) = \sum \limits_{1\leq i< j\leq  n} G(d (z_i, z_j, \varphi_i, \varphi_j) )
\label{initial_potential_1}
\end{equation}
where $G:D\to \mathbb{R}$, $D\subseteq [0, \infty)$, is some given smooth function. 
%
%
%
%
%
%
Throughout the paper, unless otherwise stated, $G$  is assumed to be non-constant.

The Euler-Lagrange equations of motion associated to $L$ are
\begin{align}
 &m_i  \frac{d}{dt} \left[(1+ f'^2(z_i)) \dot z_i \right]= m_i \dot z_i^2 f'(z_i) f''(z_i) + m_i \dot \varphi^2 f(z_i) f'(z_i) - \frac{\partial V}{\partial z_i} 
\label{lag_eq_1}\\
 &m_i  \frac{d}{dt}\left[ f^2(z_i) \dot \varphi_i  \right]= - \frac{\partial V}{\partial \varphi_i}  \quad \quad \quad i=1,2,\ldots,n\,.
 \label{lag_eq_2}
\end{align}
or 
\begin{align}
 &  \dot z_i = v_{z_i} \label{ffirst_lag}
 \\
 &\dot v_{z_i}=-  \frac{ f''(z_i)f'(z_i)}{1+ f'^2(z_i)} v_{z_i}^2+   \frac{f(z_i) f'(z_i)}{1+ f'^2(z_i)} v_{\varphi_i}^2 - \frac{2f(z_i) f'(z_i)}{1+ f'^2(z_i)}  v_{z_i}^2- \frac{1}{m_i(1+ f'^2(z_i))}\frac{\partial V}{\partial z_i}\,,
 \label{first_lag}\\
 %
 &    \dot \varphi_i =v_{\varphi_i}    \label{ssecond_lag}\\
 & \dot v_{\varphi_i}= - \frac{2 f'(z_i)}{f(z_i)} v_{z_i} v_{\varphi_i}-\frac{1}{m_if^2(z_i)}\frac{\partial V}{\partial \varphi_i}  \quad \quad \quad i=1,2,\ldots,n\,.
  \label{second_lag}
 \end{align}

\bigskip

\begin{remark} \label{config_space}

Throughout  the paper,    we assume that  all  the vector fields  are defined on domains on which they are smooth enough for the purpose in question.
%
In particular, we  assume that configurations for which the Euler-Lagrange vector field \eqref{ffirst_lag}-\eqref{second_lag}  is not at least ${\cal C}^1$  are excluded. 

For instance, observe that  for a pair of bodies the  distance function $d$  defined by \eqref{dist_function}  is smooth at all points on its domain, except   at those for which the  bodies are  on the same parallel circle and diametrically opposite. On such a circle,  the distance between the bodies 
$\tilde d(\varphi): =d(z,z, \varphi, \varphi-\pi/2) $ behaves  as  the absolute value function $f(\varphi)=|\varphi-\pi/2|$ at $\varphi=\pi/2$ and so it is Lipschitz.  
Since  we take $G$    smooth,  the composition function $G \circ d$ is at least Lipschitz on   its domain $\tilde D:=\{(z_i, z_j, \varphi_i, \varphi_j) \in (-a, b) \times (-a, b) \times \mathcal{S}^1 \times \mathcal{S}^1 \,|\, d((z_i, z_j, \varphi_i, \varphi_j)) \in D \}$ and so  the local existence and uniqueness  of the ODE solutions of the Euler-Lagrange equations is  guaranteed everywhere on the $\tilde D$. However,  part of our study requires better smoothness of the vector field (in particular the analysis of the relative equilibria bifurcations).Thus, without further notice,   we consider that all  the vector fields  appearing in the paper are sufficiently smooth.
\end{remark}
\begin{definition}\label{generalized}
The generalized n-body problem on a surface of revolution consists in the dynamics induced by the Lagrangian \eqref{Lagrangian} with a potential of the form 
\eqref{initial_potential_1}.
 \end{definition}

\begin{remark}
\label{free_motion}
If the potential is  constant   the Euler-Lagrange  \eqref{ffirst_lag}-\eqref{second_lag} 
become the geodesic equations on $\mathbb{M}$ for $n$ free mass  points. 

%
\end{remark}

\begin{proposition}[Parallel geodesic circles are invariant manifolds]
\label{par-Inv_man}
Assume that $z=z_c$ is an isolated critical point of $f$ and so $z=z_c$ is a parallel geodesic circle.  If  the bodies have their  initial positions  on the parallel geodesic circle $z=z_c$ and  their initial  velocities  tangent to that circle,  then their motion, on its domain of existence, will remain on that geodesic circle for all times. 

\end{proposition}

\noindent
Proof: 
%
%
%
At the initial time $t_0$ we have  the initial positions 
 $\q_{i0} = \left(z_i(t_0)\,, \varphi_i(t_0))\right) = (z_c\,,\varphi_{i0})$ and 
null components of the  velocities along the parallel circle $z=z_c$, i.e.
\[
v_{z_i}(t_0)=0 \quad \quad \text{for all} \quad i\,.
\]
Thus at the initial time the ODE  system \eqref{ffirst_lag}-\eqref{first_lag}  reads:
\begin{align}
 &  \dot z_i (t_0)= 0
 \nonumber \\
 &\dot v_{z_i}(t_0)= -\frac{1}{m_i}\frac{\partial V}{\partial z_i}\Big|_{\q_i=\q_{i0}}\,.
 \end{align}
%
%
%
%
Thus it is sufficient to show that 
$\displaystyle{
\frac{\partial V}{\partial z_i}\Big|_{\q_i=\q_{i0}}=0\,.
}$
The key observation is that since at the initial time  all distances $d\left( \q_{i0}\,,  \q_{j0} \right)$  are arcs of the (same) parallel geodesic circle,   the tangent vectors to $d\left( \q_{i0}\,,  \q_{j0} \right)$  are tangent to that parallel geodesic circle. Equivalently, we have that  at the initial time, the component along the meridian $ \varphi =\varphi_{i0}$ of the tangent vector to  the arc $d\left( \q_{i0}\,,  \q_{j0} \right)$  is null, i.e.
\[
\frac{\partial d_{ij} (\q_i\, \q_j)}{ \partial  z_i} \Big|_{t=t_0} =0.
 \]
Since
\begin{align}
\frac{\partial V(\q) }{\partial z_i} \Big|_{\q_i=\q_{i0}}= 
\sum\limits_{j\neq i\,,j=1}^n\,g' \,\left(d (\q_{i0}, \q_{j0}) \right) 
 \frac{ \partial d (\q_i\, \q_j)}{ \partial  z_i} \Big|_{t=t_0}=0
\end{align}
the conclusion follows.$\,\square$

\bigskip
Analogously, we have

\begin{proposition}[Meridians are invariant manifolds]
If  the bodies have their  initial positions  on a meridian with  their initial  velocities  tangent to that meridian,  then their motion, on its domain of existence, will remain on that meridian circle for all times.  

\end{proposition}

\subsection{Hamiltonian formulation and momentum conservation}

By applying the Legendre transform to \eqref{Lagrangian} we  find the corresponding Hamiltonian $H: T^*\mathbb{Q} \to \mathbb{R}$
\begin{equation}
H(\q,  \p) = K_\q(\dot \q) +V(\q)
\label{Ham_form}
\end{equation}
with the kinetic energy
\begin{equation}
 K_\q(\p) = \frac{1}{2}
 \sum\limits_{i=1}^n  \frac{1}{2m_i}
\left[p_{z_i}\, p_{\varphi_i} \right]
\left[
\begin{array}{cc}
\frac{1}{1+ f'^2(z_i)} & 0 \\
0 & \frac{1}{f^2(z_i)}
\end{array}
\right]
 \left[
\begin{array}{c}
 p_{z_i} \\
p_{\varphi_i}
\end{array}
\right]
\end{equation}
where   $\p_i:=(p_{z_i}\, p_{\varphi_i}) \in \mathbb{R}^2$ and $\p:=(\p_1, \p_2, \ldots \p_n)$.
The equations of motion are
\begin{align}
&\dot z_i = \frac{p_{z_i}}{m_i(1+ f'^2(z_i))}\,,\quad \quad \quad \dot p_{z_i} = \frac{f'(z_i)f''(z_i)}{m_i(1+ f'^2(z_i))^2}p_{z_i}^2+\frac{f'(z_i)}{m_if^3(z_i)}p_{\varphi_i}^2-\frac{\partial V}{\partial z_i}\,, \label{eq_ham_1}\\
&\dot \varphi_i = \frac{p_{\varphi_i}}{m_i f^2(z_i)}\,, \quad \quad  \quad \quad \quad \dot p_{\varphi_i} = -\frac{\partial V}{\partial \varphi_i}\,.  \label{eq_ham_2}
\end{align}

\bigskip
\begin{remark}
\label{Ham_free_motion}
If the potential is constant  we retrieve the Hamiltonian formulation of the dynamics of $n$ free mass  points on $\mathbb{M}$. The trajectories $\left(z_i(t)\,, \varphi_i(t) \right)$ describe geodesics on the surface. 
\end{remark}

\bigskip
\noindent
The rotation group $SO(2)$ acts isometrically on $\mathbb{M}$ by $(R_\alpha, (z, \varphi)) \to (z, \varphi + \alpha)\,,$ where $R_\alpha$ represents a rotation of angle $\alpha$ and $(z, \varphi) \in \mathbb{M}\,.$ 
Further,  $SO(2)$ acts on $T\mathbb{M}$ by 
\begin{equation}
(R_\alpha, (z, \varphi, v_z, v_{\varphi})) \to  (z, \varphi +\alpha, v_z, v_{\varphi}),
\label{action_TS}
\end{equation}
and on  $T^*\mathbb{M}$ by \begin{equation}
(R_\alpha, (z, \varphi, p_z, p_{\varphi})) \to  (z, \varphi + \alpha, p_z, p_{\varphi})\,.\label{action_T_Star_S}
\end{equation}
The associated infinitesimal generator vector field is 
\begin{equation}
\omega\cdot (z,\varphi) =\frac{d}{dt}\Big|_{t=0} \exp(t\omega) (z, \varphi) = (0\,, \omega)\,,\quad \quad \omega \in so(2)\,, \,\,(z,\varphi) \in \mathbb{M}\,.
\label{generator}
\end{equation}
%
%
The rotational actions on $\mathbb{M},$ $T\mathbb{M}$ and $T^*\mathbb{M}$  extend naturally to  diagonal acting on $\mathbb{M}^n$, $T\mathbb{M}^n$ and $T^*\mathbb{M}^n$.
%
%
%
For instance, the infinitesimal generator corresponding to the rotation group action on $T\mathbb{M}^n$ is given by
\begin{align}
\omega \cdot \q &=\omega \cdot (\q_1,\q_2, \ldots ,\q_n) \nonumber  \\
&= \omega \cdot ((z_1, \varphi_1), (z_2, \varphi_2), \ldots ,(z_n, \varphi_n))  \nonumber  \\
& =  (  \omega \cdot (z_1, \varphi_1),  \omega \cdot (z_2, \varphi_2), \ldots , \omega \cdot (z_n, \varphi_n))
 = ((0\,, \omega)\,, (0\,, \omega) \ldots, (0\,, \omega))\,.
 \label{generator-gen}
 \end{align}
%
Associated to the rotational group action on $T^*\mathbb{M}^n$ is the (angular) momentum map  $J: T^*\mathbb{M}^n \to so^*(2)$ which may be calculated using the co-tangent bundle formula
$ \left< J(\q,\p)\,,\omega \right>= \left<\p\,, \omega \cdot \q \right>$ for all $\omega \in so(2)$ (see \cite{HSS09}). We obtain
\begin{align}
J(\q, \p)=J\left((z_1, \varphi_1 , p_{z_1}, p_{\varphi_1}), (z_2, \varphi_2 , p_{z_2}, p_{\varphi_2})\,,\ldots, (z_n, \varphi_n, p_{z_n}, p_{\varphi_n}) \right)= p_{\varphi_1} +p_{\varphi_2}+\ldots +p_{\varphi_n}   \,.
\nonumber
\end{align}

\bigskip
Since  the Hamiltonian \eqref{Ham_form} is rotationally invariant under the diagonal action of the  rotation group  on  $T^*\mathbb{Q},$ %
by Noether theorem, the  momentum $J(\q, \p)$ is constant along any solution $\displaystyle{\left(\q(t),\p(t) \right)}$, and so
\begin{align}
J(\q(t),\p(t)) = p_{\varphi_1}(t) +p_{\varphi_2}(t)+\ldots +p_{\varphi_n}(t) = const.:= \mu\,.
\label{mom-map}
\end{align}

The effect of rotations on the bodies is given by the (locked) \textit{inertia tensor}, denoted $\mathbb{I},$ which maps  each point $\q$ to  a linear application $\mathbb{I}(\q):   so(2) \to so^*(2)$
\begin{align}
\left<\mathbb{I}(\q)(\omega)\,,\,\eta\right>_{so(2)} := \ll \omega \cdot \q\,, \eta\cdot \q  \gg
\end{align}
where $\left< \cdot\,, \cdot \right>_{so(2)}$ is the pairing between $so^(2)$ and its dual $so^*(2)$ (in our case, just number multiplication), and $\ll \cdot \,, \cdot\gg$ is the metric on $T\mathbb{M}^n.$ Since $\text{dim} \,so(2)=1,$ the inertia tensor $\mathbb{I}$   is  a scalar  and is in fact the analogue of the usual moment of inertia associated to the action of the rotation group on the plane. 
Using \eqref{generator-gen} and \eqref{metric} we obtain that the moment of inertia for  the motion of $n$ mass points on a surface of revolution is 
 \begin{align}
\mathbb{I}(\q)= \mathbb{I}((z_1, \varphi_1), (z_2, \varphi_2), \ldots ,(z_n, \varphi_n))=\sum \limits_{i=1}^n m_if^2(z_i)\,.
\label{moment_inertia}
\end{align}

\subsection{Relative equilibria}

For a Hamiltonian system with continuous  (Lie) symmetry, relative equilibria (RE) are solutions which are also (one-parameter) orbits of the  symmetry group (\cite{Ma92}). In our case,  taking into account the action \eqref{action_T_Star_S}, its diagonal extension to $\mathbb{M}^n$ and the lift to $T^*\mathbb{M}^n \supset T^*\mathbb{Q}$\,, these solutions must be of the form
\begin{equation}
z_i(t)=const.=z_{i0} \,,\quad 
 \varphi_i(t)=  \omega\, t + \varphi_{i0} \,,\quad p_{z_i}(t)= const. =p_{z_{i0}}\,,\quad  p_{\varphi_i}(t) = const.=p_{\varphi_{i0}} 
 \label{def_RE}
\end{equation}
for some \textit{group velocity}  $\omega \in so(2)\simeq \mathbb{R}$  and  some suitable \textit{base point}  $(\q_0\,, \p_0) \in T^*\mathbb{Q}$, $\q_0 = (\q_{10}\,, \q_{20}\,, \ldots\,, \q_{n0})$\,, $\p_0 = (\p_{10}\,, \p_{20}\,, \ldots\,, \p_{n0})$\,, 
$\q_{i0} = (z_{i0}\,, \varphi_{i0})$\,, $\p_{i0} = (p_{z_{i0}}\,,p_{\varphi_{i0}})\,.$
The base points $\q_{i0}=(z_{i0}\,,\varphi_{i0})$ are found 
 as critical points of the \textit{augmented potential} (\cite{Ma92})
%
\begin{equation}
V_{\omega}\left((z_1, \varphi_1), (z_2, \varphi_2), \ldots ,(z_n, \varphi_n) \right)=  -  \frac{\sum \limits_{i=1}^nm_if^2(z_i) \omega^2}{2}  + V\left( (z_1, \varphi_1), (z_2, \varphi_2), \ldots ,(z_n, \varphi_n)\right)\,
\end{equation}
that is, solutions of
\begin{align}
- \omega^2 m_i f(z_i)  f'(z_i)   + \frac{\partial V}{\partial z_i}&=0 \label{ore_1} \\
\frac{\partial V}{\partial \varphi_i} &=0 \label{ore_2} 
\end{align}
whereas 
\[
p_{z_{i0}}=0\,,\quad\quad p_{\varphi_{i0}}= m_i f^2(z_{i0})\omega\,.
\]
Alternatively, one may  determine the  base points  as critical points of the \textit{amended potential} 
\begin{equation}
V_{\mu}\left((z_1, \varphi_1), (z_2, \varphi_2), \ldots ,(z_n, \varphi_n) \right)=  \frac{\mu^2}{2 \sum \limits_{i=1}^n m_i f^2(z_i)} + V\left( (z_1, \varphi_1), (z_2, \varphi_2), \ldots ,(z_n, \varphi_n)\right)\,.
\end{equation}
where $\mu$ is  (a fixed value of) the angular momentum.
Thus one  needs to solve
\begin{align}
- \mu^2 \frac{m_i f(z_i)  f'(z_i)}{\left(\sum \limits_{i=1}^n m_i r^2(z_i)\right)^2}   + \frac{\partial V}{\partial z_i}&=0 \label{re_1} \\
\frac{\partial V}{\partial \varphi_i} &=0 \label{re_2} 
\end{align}
The relationship between the group velocity $\omega$ and the angular momentum $\mu$ is
\begin{align}
\mu=\mathbb{I}(\q) \omega = \left(\sum \limits_{i=1}^n m_i r^2(z_i)\right) \omega
\end{align}
and can be retrieved following a general  geometric mechanics context (see \cite{Ma92})\,.
%
 %
 %
 %
We also note that a  RE with zero $\omega=0$, or equivalently, with $\mu=0,$ is in fact an equilibrium.

\begin{definition}
A solution  is called Lagrangian if, at every time t, the masses form a polyhedron   that is orthogonal to the z axis.
\end{definition}

\begin{proposition}
If there is  an equilibrium so that the points lie on a parallel geodesic circle 
then 
 for every nonzero group (angular) velocity $\omega \neq 0 $ there is a  RE with the same configuration  that rotates along the parallel geodesic circle.
\end{proposition}

\noindent Proof: Let $z=z_{c}$ be a parallel geodesic circle. Assume there is an equilibrium $(z_i=z_{c}\,, \varphi_{i}=\varphi_{i0})$, $i=1,2\ldots n$ and so
\begin{align}
 \frac{\partial V}{\partial z_i} \Big|_{z_i=z_{cr}\,, \varphi_{i}=\varphi_{i0}} =\frac{\partial V}{\partial \varphi_i} \Big|_{z_i=z_{c}\,, \varphi_{i}=\varphi_{i0}}=0 \nonumber\,. 
\end{align}
Since $f'(z_{c})=0$ and RE are given by \eqref{ore_1}-\eqref{ore_2}, the conclusion follows.

\noindent
$\square$






\subsection{Motion on a cylinder and the ``vertical linear momentum" integral}

If $\mathbb{M}$ is a cylinder, i.e. $f'(z)=0$ and $f(z)=const.=: f_0$ for all $z$, then the Hamiltonian \eqref{Ham_form} becomes
\begin{equation}
H(\q,  \p) = \frac{1}{2}
 \sum\limits_{i=1}^n  \frac{1}{2m_i}
\left[p_{z_i}\, p_{\varphi_i} \right]
\left[
\begin{array}{cc}
1 & 0 \\
0 & \frac{1}{f_0^2} 
\end{array}
\right]
 \left[
\begin{array}{c}
 p_{z_i} \\
p_{\varphi_i}
\end{array}
\right]
+ V\left(d (\q_i, \q_j)\right)
\,.
\label{Ham_cyl}
\end{equation}
and the equations of motion are
\begin{align}
&\dot z_i =  \frac{p_{z_i}}{m_i}\,, \quad \quad \quad \, \dot p_{z_i} =- \frac{\partial V\left(d (\q_i, \q_j) \right)} { \partial z_i} 
\label{eq_z_cyl} \\
&\dot \varphi_i =  \frac{p_{\varphi_i}}{m_i f_0^2}\,, \quad \quad \dot p_{\varphi_i} =- \frac{\partial V\left(d (\q_i, \q_j)\right)}{ \partial \varphi_i}
\end{align}

%
%
%
%
On a cylinder the geodesic distance between two points $(z_i, \varphi_i)$ and $(z_j, \varphi_j)$ is invariant to translations along the $z$ direction:
\[
d (\q_i, \q_j) =  d \left((z_i, \varphi_i), (z_j, \varphi_j) \right)=d(z_i-z_j, \varphi_i-\varphi_j)\,. 
\]
It follows that the Hamiltonian  is invariant to translations along the $z-$axis, that is it is invariant under the diagonal action of $\mathbb{R}$ on $T^*\mathbb{M}^n$, where the action on each component is 
\[
(x, (z_i, \varphi_i, p_{z_i}, p_{\varphi_i} )) \to (z_i +x, \varphi_i, p_{z_i}, p_{\varphi_i} )
\,, \quad \,\, x \in \mathbb{R}\,,\,\, (z_i, \varphi_i)\in \mathbb{M}\,,\,\,\, (p_z, p_{\varphi}) \in T^*_{(z_i, \varphi_i)}\mathbb{M}\,.
\]
This symmetry  leads to the conservation of the ``vertical linear momentum"
\begin{align}
L(t):=p_{z_1}(t) + p_{z_2}(t) + \ldots p_{z_n}(t) =const.=: k=L(t_0)
\label{vertical}
\end{align}
which can be found by either calculating the corresponding momentum map and  using Noether's theorem, or by a direct guess and verification. Further, summing the  equations for $z_i$ from  \eqref{eq_z_cyl},  integrating, and taking into account \eqref{vertical}, we obtain the \textit{vertical centre of mass} integral
\begin{align}
M(t):=m_1 z_1(t) + m_2 z_2(t)  + \ldots m_n z_n(t)   = k t + k_0 
\end{align}
where $k_0 =M(t_0) -kt_0\,.$

\section{Saari's conjecture on surfaces of revolution}

From   formula \eqref{moment_inertia} we deduce that  the moment of inertia on surfaces of revolution  $\mathbb{I}$ is constant when all points are on the same parallel circle.   
In particular, since  by Proposition \ref{par-Inv_man} parallel  geodesic circles are invariant manifolds, we obtain the following:
\begin{proposition}
Any motion of the $n$-bodies on a geodesic parallel circle  has constant moment of inertia.
\end{proposition}
\begin{corollary}
The generalization of Saari's Conjecture to $n$-body problems on surfaces of revolution with at least one geodesic  parallel circle is not true.
\end{corollary}
In particular we settle the   the conjecture posed on surfaces of constant positive curvature  as presented  by Diacu \& al. in \cite{Diacu2012}:

\begin{corollary}
The generalization of Saari's Conjecture to $n$-body problems on a 2-spheres  is not true.
\end{corollary}
Also, we have:
\begin{corollary}
The generalization of Saari's Conjecture to $n$-body problems on a cylinder  is not true.
\end{corollary}

%
%


\section{Equal masses and homographic motions}

\subsection{Homographic motions as a symplectic invariant manifold}

\begin{definition}
A solution of the generalized $n$-body problem on a surface of revolution   is \underline{homographic} if it is of the form 
\begin{align}
z_k(t) =z(t)\,, \quad \varphi_k(t) = \varphi(t) + \theta_k\,,\quad \quad \,k=1,2, \ldots, n\,.
\label{hom-inv-gen}
\end{align}
for some functions $z(t)$ and $\varphi(t)$ and some constant angles $\theta_k \in [0, 2\pi)$ . 
\end{definition}
In other words, the particles are all in a plane orthogonal to the $z$ axis and move  simultaneously along the given surface of revolution  while keeping a self-similar shape in the ambient $\mathbb{R}^3$ space at all times. 
When the dynamics is given in Hamiltonian formulation, a solution of the Hamiltonian system \eqref{eq_ham_1}-\eqref{eq_ham_2} is called homographic if it is of the form 
\begin{align}
z_k(t)=z(t)\,,  \,\,\,\varphi_k(t) = \varphi(t) + \theta_k\,, \,\,\,p_{z_k}(t)=p_z(t)\,,  \,\,\,p_{\varphi_k}(t)= p_{\varphi}(t)\,,  \quad \quad \,k=1,2, \ldots, n\,.
\label{L_homographic_Ham}
\end{align}
%

If they exist, homographic solutions are parametrized by the their (constant) angular momentum $p_{\varphi}.$ Specifically, let $z(t)\,, \varphi(t)\,,p_z(t)\,,p_{\varphi}(t)$ be a homographic solution.  Using the total angular momentum conservation \eqref{mom-map} and since all homographic momenta \eqref{L_homographic_Ham} are equal  we have
\begin{align}
\sum\limits_{k=1}^n p_{\varphi_k}(t)=n p_{\varphi}(t)=\mu \quad \text{and so} \quad  
p_{\varphi}(t) = \mu/n=:c\,.
\label{moment_homographic}
\end{align}
Thus to every   homographic solution corresponds a value of the angular momentum $c.$ To  zero momentum $c=0$  correspond $\text{homothetic}$ motions, that is solutions for which  the particles  move synchronously  along meridians. For values $c\neq0$ we obtain purely homographic solutions; the particles move rotate while "sliding" at an unison on the surface. Homographic motions such that $z(t)= const.=:z_0$ for all $t$ are  RE; this can be seen from the RE equations \eqref{def_RE}.


\subsection{Discrete reduction and the equal mass homographic invariant manifold}

\bigskip
When the masses are equal, homographic motions form an invariant manifold for which  the points remain at the vertices of a regular $n$-gon  and move simultaneously on  paths along $\mathbb{Q}$. %
We retrieve this  manifold   by applying the method of  
\textit{Discrete reduction} (see \cite{Ma92, Montaldi98}) that we  briefly recall below.

\bigskip
\noindent
Let $\Sigma$ be a discrete  group act on a cotangent bundle $T^*Q$. Its fixed point set $\text{Fix}\,(\Sigma, T^*Q)$ is defined by:
\begin{equation}
\text{Fix}\,(\Sigma, T^*Q):= \{(q,p) \in T^*Q\,|\, g(q,p)=(q,p)\,\,\,\forall\, \sigma \in \Sigma\}\,.
 \end{equation}
If a Hamiltonian $H: T^*Q \to \mathbb{R}$  is $\Sigma$-invariant and the symplectic structure is preserved under the $\Sigma$-action, then $\text{Fix}\,(\Sigma, T^*Q)$ is an invariant manifold for the dynamics of $H$. If in addition, $\Sigma$ is compact, then $\text{Fix}\,(\Sigma, T^*Q)$ is a symplectic invariant submanifold, and the restrictions of the Hamiltonian and the symplectic form to $\text{Fix}\,(\Sigma, T^*Q)$ is a co-tangent bundle Hamiltonian system which coincides with the restriction of the given Hamiltonian system. Moreover, if the symplectic structure and $H$ are invariant under the action of a Lie group $G$ giving rise to an equivariant momentum map, and  $\Sigma$ acts on $G$ in such a way that the actions  of $G$ an $\Sigma$ are compatible then $\text{Fix}\,(\Sigma, T^*Q)$ is a co-tangent bundle Hamiltonian system with $G$ symmetry. 
%
 Also, by  \textit{Palais' Principle of Criticality} (\cite{Palais68}) any equilibrium or RE in $\text{Fix}\,(\Sigma, T^*Q)$ is also an equilibrium or a RE, respectively, in the full $T^*Q$ phase space.

 \bigskip
Returning to the $n$-body problem on surface of revolution,  assume that all masses are equal, i.e.,  $m_1=m_2=\ldots=m_n$. 
Now consider  the cyclic group $C_n$ generated by  counterclockwise rotations $R_{2\pi/n}$ of angle $2\pi/n$ about the vertical; this group is isomorphic to $(\mathbb{Z}_n, +)$.  Also, consider  the subgroup of permutations  $S_n$ of the set $\{1,2,\ldots, n\}$  generated by a shift to the right by one unit of the elements, which we denote $\sigma$.

 \bigskip
\noindent
The product group $S_n \times C_n$  acts on $\mathbb{M}^n$ by first rotating (all) the points by a multiple of $2\pi/n$, and  then relabelling the masses. A generator for this action is 
\begin{align}
& \left((\sigma\,, {\cal R}_{2\pi/n})\,, \left( (z_1, \varphi_1),  (z_2, \varphi_2),\ldots, (z_n,  \varphi_n) \right) \right) 
 \to (\sigma\,,{\cal R}_{2\pi/n}) \cdot \left( (z_1, \varphi_1),  (z_2, \varphi_2),\ldots, (z_n,  \varphi_n) \right)    \nonumber  \\
& \hspace{4.7cm} :=\left( (z_n,  \varphi_n+2\pi/n), (z_1, \varphi_1+2\pi/n),  \ldots, (z_{n-1},  \varphi_{n-1}+2\pi/n) \right)   
\end{align}
Further, $S_n \times C_n$  acts on $T^*\mathbb{M}^n$ by co-tangent lift
\begin{align}
&\left( (\sigma\,,{\cal R}_{2\pi/n})\,, \left((z_1, \varphi_1, p_{z_1}, p_{\varphi_1}),  (z_2, \varphi_2, p_{z_2}, p_{\varphi_2}),\ldots, (z_n, \varphi_n, p_{z_n}, p_{\varphi_n}) \right)  \right) \nonumber\\
 &\to (\sigma\,,{\cal R}_{2\pi/n}) \cdot \left((z_1, \varphi_1, p_{z_1}, p_{\varphi_1}),  (z_2, \varphi_2, p_{z_2}, p_{\varphi_2}),\ldots, (z_n, \varphi_n, p_{z_n}, p_{\varphi_n}) \right)
\\
&:=\left( (z_n,  \varphi_n+2\pi/n \,, p_{z_n}, p_{\varphi_n}), (z_1, \varphi_1+2\pi/n \,, p_{z_1}, p_{\varphi_1} ),  \ldots, (z_{n-1},  \varphi_{n-1}+2\pi/n \,, p_{z_{n-1}}, p_{\varphi_{n-1}} ) \right) 
\end{align}
It follows that the fixed point  set $\text{Fix}(S_n \times C_n)$  os %
\begin{align}
& \text{Fix}(S_n \times C_n)= \left\{(z_1,p_{z_1}, \varphi_1, p_{\varphi_1}, z_2,p_{z_2}, \varphi_2, \, p_{\varphi_2},,\ldots, z_n, p_{z_n}, \varphi_n, p_{\varphi_n}) \, | \,  z_i=z_j\,,\, p_{z_i}=p_{z_j}\,,\right. \nonumber \\
&\quad \quad \quad \left. p_{\varphi_{i}}=p_{\varphi_{j}}\,\,\, \forall\, i\,,j\,\,\text{and}\,\,\, \varphi_{i+1}- \varphi_{i}=2\pi/n\,,\,\,i=1,2\ldots (n-1)\,, \varphi_{1}- \varphi_{n}=2\pi/n \,\right\}, 
\label{inv_homogr}
\end{align}
which coincides to the  set of homographic solutions \eqref{L_homographic_Ham}  with  constant angles $\theta_k=2\pi/n$.
%
%
%
%
%
By the Discrete reduction method, we  have:
%
%
%
\begin{proposition}\label{lema_1}
The generalized n-body problem on a surface of revolution with equal masses admits  an  invariant manifold that is homographic with the particles forming a regular $n$-gon at al times.
If the problem is modeled in Hamiltonian formulation, this invariant manifold is symplectic and on it the dynamics is given by a co-tangent bundle Hamiltonian system.
\end{proposition}

\begin{remark}
To ease the reading of the paper, we avoid checking the compatibility condition of the actions of $S_n \times C_n$ and $SO(2)$ on $T^*\mathbb{M}$ as given by Assumption $1_Q$  in Marsden's  book \cite{Ma92}, pp.155,  which would guarantee that the homographic Hamiltonian system has $SO(2)$ symmetry. Instead we will just write the homographic Hamiltonian and observe  that it is rotationally symmetric. 
\end{remark}

Thus, when the masses are equal, homographic solutions are of the form
\begin{align}
z_k(t) =z(t)\,, \quad \varphi_k(t) = \phi(t) + \frac{2k\pi}{n}\,,\quad \quad \,k=1,2, \ldots, n\,.
\label{hom-inv}
\end{align}
Equivalently,  if the masses are equal, a solution of the Hamiltonian system \eqref{eq_ham_1}-\eqref{eq_ham_2} is  homographic if it is of the form 
\begin{align}
z_k(t)=z(t)\,,  \,\,\,\varphi_k(t) = \phi(t) + \frac{2k\pi}{n}\,, \,\,\,p_{z_k}(t)=p_z(t)\,,  \,\,\,p_{\varphi_k}(t)= p_{\varphi}(t)\,,  \quad \quad \,k=1,2, \ldots, n\,.
\label{homographic_Ham}
\end{align}

\begin{definition}
An equal mass  homographic trajectory is a curve on $\mathbb{Q}$ given by the  configuration $\left(z(t), \phi(t)\right)$  of a homographic solution \eqref{homographic_Ham} defined above. 

\end{definition}

From now on we assume that all  bodies have equal mass, which without loosing generality, we take to be unity.  Also, to simplify denominations, \textit{in what follows we'll refer to equal mass  homographic dynamics, solutions,  trajectories, etc. as homographic dynamics, solutions,  trajectories, etc.} However, given appropriate specifications for the masses, there are other situations in which  homographic dynamics is present. For instance, it is easy to see that for a three body problem, if two masses are equal, solutions for which the bodies  are all in a plane orthogonal to the axis of revolution and form isosceles triangles  at all times are homographic; in this case, the discrete symmetry group is $S_2 \times \mathbb{Z}_2,$ (acting as a reversal of the equal masses followed by their  relabelling).

\subsection{Dynamics. The generalized Clairaut relation}

The dynamics on the homographic invariant manifold is given by  the Hamiltonian 
\begin{align}
&\tilde H: (-a\,,b) \times {\mathcal{S}}^1 \times \mathbb{R} \times \mathbb{R} \to \mathbb{R} \nonumber\\
&\tilde  H(z,  \varphi\,, p_z, p_{\varphi}) = 
\frac{n}{2}
\left[
\begin{array}{cc}
p_{z} & p_{\varphi} 
\end{array}
\right]
\left[
\begin{array}{cc}
\frac{1}{1+ f'^2(z)} & 0 \\
0 & \frac{1}{f^2(z)}
\end{array}
\right]
 \left[
\begin{array}{c}
 p_{z} \\
p_{\varphi}
\end{array}
\right]+ W(z)
\label{ham_homo}
\end{align}
%
%
where
\begin{align}
W\left(z \right) := \sum\limits_{1\leq i<j\leq n}
G\left(
\frac{2\pi(j-i)}{n}f(z)
\right)\,. 
\label{pot_homo}
\end{align}
The equation of motion are:
\begin{align}
&\dot z = \frac{n\,p_{z}}{1+f'^{2}(z)} 
 \label{eq_hom_orig_1} \\
&\dot p_{z} =  - nf'(z)\left( \frac{f''(z) \,p_z^2}{\left(  1+f'^{2}(z) \right)^2} - \frac{p^2_{\varphi}}
{f^3(z)} +
 \sum \limits_{k=1}^{n-1} \frac{2(n-k)k\pi}{n}G'\left(\frac{2k\pi}{n}\, f(z)  \right) \right) \label{eq_hom_orig_2}  \\
& \dot \varphi =  \frac{np_{\varphi}}{f^2(z(t))}
 \label{eq_hom_orig_3} \\
& \dot p_{\varphi}  = 0\,.
   \label{eq_hom_orig_4}
\end{align}
The motion  of body $k$, $k=1,2,\ldots, n$ is a solution of  the initial value problem given by the system above and initial conditions of the form
\begin{equation}
z_k(t_0)=z_0\,,\quad  p_{z_k}(t_0)=p_{z,0}\,,\quad  \varphi_k(t_0)=\phi_(t)+\frac{2k\pi}{n}\,,\quad  p_{\varphi_k}(t_0) = p_{\varphi,0}\,.
\end{equation}
\begin{remark}
\label{hom_free_motion}
For a constant potential (i.e., $G\equiv const.$ ) we retrieve the Hamiltonian formulation for the homographic dynamics of $n$ free mass  points on $\mathbb{M}$. 
The bodies   are  situated on a plane perpendicular to the axis of revolution at all times with each trajectory $\left(z_k(t)\,, \varphi_k(t) \right) = \left(z_0\,, \phi_(t)+\frac{2k\pi}{n} \right)$ describing a geodesics on the surface. 
%
\end{remark}
The conservation of energy reads
\[
\tilde  H\left(z(t), \varphi(t), p_{z}(t), p_{\varphi}(t)  \right)= const.=: h\,.
\]
%
%
%
The angular momentum conservation
\begin{align}
p_{\varphi}(t)=const.=: c\,.
\label{total_ang_momentum_map}
\end{align}
is retrieved immediately from equation \eqref{eq_hom_orig_4}. (Alternatively, one can use the total angular momentum conservation given by  \eqref{mom-map} specified to the context of homographic motions).
Note that 
\begin{equation}
n c = \mu
\label{momentum _equal}
\end{equation}
where  $\mu$ is the total angular momentum  given by \eqref{mom-map}.

\bigskip
\begin{proposition}(Clairaut relation for homographic motions)
Let $\left(z(t) , \varphi(t)\right)$ be a homographic  trajectory, and denote by $\theta(t)\in (0, \pi/2)$ the angle of the trajectory with the parallel circle $z(t).$ Then
 the product of the  radious' trajectory (i.e. $f(z(t))$), speed, and $\cos$ of the angle  $\theta(t)\in (0, \pi/2)$    is constant, that is
 \begin{equation}
 \left(  \sqrt{
 \left(1+f'^2\left(z(t)\right) \right)
  \left(\frac{dz}{dt}\right)^2
  +f^2 \left(z(t)\right) 
  \left(
  \frac{d\varphi}{dt} \right)^2
   }
   \right)
   f\left(z(t)\right) 
   \cos
   \left(\theta(t) 
   \right) = const. = nc=\mu\,.
 \label{Clairaut}
 \end{equation}
 \end{proposition}
 
 \noindent
 Proof. 
 Let  $\left(z(t)\,, \varphi(t)\right)$ be a solutions' trajectory on $\mathbb{Q}$. 
 Substituting $p_{\varphi} (t)=c$ into \eqref{eq_hom_orig_1} we write
 \begin{equation}
f^2\left(z(t)\right)\frac{d\varphi}{dt} =  nc\,.
\label{ang_Clairaut}
  \end{equation}
Given the $(z, \varphi) \to \x(z, \varphi)$ parametrization in equation \eqref{param_M}, we have
\begin{align*}
 \x_z = \left(f'(z)\cos\varphi\,,   f'(z)\sin\varphi\,, 1\right), \quad \x_\varphi = \left( -f(z)\sin\varphi\,,   f(z)\cos \varphi\,, 0\right)\,. 
\end{align*}
Thus
 \begin{equation}
 \cos(\theta(t)) = \frac{\left< \x_z \left(\frac{dz}{dt}\right)^2 + \x_\varphi   \left(\frac{d\varphi}{dt} \right)^2 \,,  \x_\varphi \right>}
 {\left\|\x_z \left(\frac{dz}{dt}\right)^2 + \x_\varphi  \left(\frac{d\varphi}{dt} \right)^2 \right\| \| \x_\varphi \|} 
  \end{equation}
 and so
 \begin{equation}
  \cos(\theta(t))=\frac{ f\left(z(t)\right)  \frac{d\varphi}{dt}}{\left\|\x_z \left(\frac{dz}{dt}\right)^2 + \x_\varphi  \left(\frac{d\varphi}{dt} \right)^2 \right\|}\,.
 \label{cos_theta}
 \end{equation}
Multiplying the relation above by $f \left( z(t) \right)$ and using \eqref{ang_Clairaut} the conclusion follows. $\square$
 
 
 \begin{remark}
 If the potential is constant (i.e. $G\equiv const.$ ), the system \eqref{eq_hom_orig_1} - \eqref{eq_hom_orig_4} describe  geodesic motion (in Hamiltonian formulation) and  the homographic trajectories   are   the surfaces' geodesics. In particular, since the speed along any geodesic is constant,  in this case the generalized Clairaut relation  \eqref{Clairaut} above becomes the well-known Clairaut relation   on a surface of revolution. 
 \end{remark}

\begin{remark}
The generalized Clairaut relation \eqref{Clairaut}  expresses the fact that when the potential is non-zero, one cannot find a parametrization for which all homographic motions have unit speed.
 \end{remark}

 \begin{remark}
 The generalized Clairaut relation \eqref{Clairaut}  is  a consequence of  the  angular momentum conservation associated to any rotational-invariant two-degrees of freedom system defined on a surface of revolution.   
 \end{remark}

 \bigskip
Substituting $p_{\varphi}(t)= c$ into  the Hamiltonian $\tilde H$ we  reduce  the dynamics to a one-degree of freedom (and thus integrable) system given by
\begin{align}
\tilde H_{\text{red}}(z, p_z)=  \frac{n}{2}\,
\frac{p_z^2}{1+ f'^2(z)}  
  + W_c(z)\,,
\label{hh_red:1}
\end{align}
where $W_c(z)$ is the \textit{amended potential}
\begin{align}
 W_c(z):= \frac{n}{2} \frac{c^2}{f^2(z)} + W(z)=\frac{n}{2} \frac{c^2}{f^2(z)} +  \sum\limits_{1\leq i<j\leq n}
G\left(
\frac{2\pi(j-i)}{n}f(z)
\right)\,.
\label{am_red:1}
\end{align}
The time evolution of the angle $\varphi$ is given by \eqref{eq_hom_orig_3} 
and thus,  once a solution of the reduced system $(z(t), p_z(t))$ is found, $\varphi(t)$ is given by the \textit{reconstruction} equation
\begin{equation}
\varphi(t)  =\varphi(t_0) +n \int\limits_{t_0}^t\ \frac{c}{f^2(z(\tau))}d\tau\,.
\label{reconstruction}
\end{equation}

\subsection{Homographic dynamics on a cylinder}

When $\mathbb{M}$ is a cylinder i.e., $f(z)=const.=: f_0$, the dynamics  reduces to a linear flow. Indeed,
in this case the Hamiltonian \eqref{hh_red:1} of the reduced dynamics on the  homographic invariant manifold is
\begin{align}
\tilde H_{red} (z, p_z) = \frac{n}{2}p_z^2 +  \frac{n}{2c^2} +  
\sum\limits_{1\leq i<j\leq n} G\left(
\frac{2\pi(j-i)}{n} f_0
\right) =  \frac{n}{2}p_z^2 +const\,,
\end{align}
and the homographic  dynamics consists  in uniform motions no matter the potential:
%
\begin{align}
&z(t)=n\,p_{0z}\,t + z_0\,,\quad \quad p_z(t)=p_{0z}\, \\
&\varphi (t)= c\, t +\varphi_0\,, \quad \quad \quad\,\,\, p_{\varphi}(t)=const.=c\,.
\end{align}
%

%


\subsection{Hill's regions and topology of the phase space}

As a   two-degrees of freedom Hamiltonian system with rotational symmetry, the dynamics on the homographic invariant manifold
 possesses  two independent integrals (the energy $\tilde H$ and angular momentum $\tilde J$) and so   is integrable.
Moreover, given the "kinetic + (amended) potential" structure of the reduced one-degree of freedom system,  a sketch  of the amended potential is sufficient for extracting a full qualitative picture of the dynamics, including the Hill's regions of motions, the topology of the phase-space and the orbit types (see \cite{Arnold78, Sm70}). This is the subject of this subsection.

\bigskip
Recall that the Hill's regions of motions of a mechanical "kinetic + (amended) potential " system at a given  fixed energy level $h$ and momentum $c$,
are defined
as the regions of allowed motion in the configuration space; the later are determined by   taking into account that the 
kinetic energy is non-negative. 
%
%
In our case the Hill's regions are
\begin{align}
{\mathcal R}_{h,c}:&=\{ (z, \varphi) \in (-a\,,b) \times {\mathbb S}^1 \,\Bigg|\, \,\, W_c(z)\leq h \} \nonumber\\
&=\left\{ (z, \varphi) \in (-a\,,b) \times {\mathbb S}^1 \,\Bigg|\, \,\, \frac{nc^2}{2f^2(z)} +  \sum\limits_{1\leq i<j\leq n}
G\left(
\frac{2\pi(j-i)}{n}f(z)
\right)
 \leq h \right\}  \,.
\end{align}
The  topological characterization of the  phase space is obtained by considering  the level sets of the \textit{energy-momentum map}
\begin{align}
&\tilde H\times \tilde J :   (-a\,,b) \times {\mathcal{S}}^1 \times \mathbb{R} \times \mathbb{R} \to \mathbb{R}  \times \mathbb{R}, \nonumber \\
&(\tilde H\times  \tilde J)(z,  \varphi\,, p_z, p_{\varphi}) = \left( \tilde H(z,  \varphi\,, p_z, p_{\varphi})\,,\tilde J(z,  \varphi\,, p_z, p_{\varphi}) \right)\,.
\end{align}
For $h \in \mathbb{R}\,, c\in \mathbb{R}$, the level sets   
 \begin{align}
 I_{h,c}:= (\tilde H \times \tilde J )^{-1}(h,c)\,.
 \label{level}
 \end{align}
are  invariant submanifolds and provide a  foliation of the phase-space.
 For instance, if we assume that the flow is complete (i.e. collisions are not possible) and $c\neq 0$, we have:
\begin{align}
I_{h\,,c} &:= ( \tilde H\times \tilde J)^{-1}(h\,,c) \\
&= \left\{ (z, \varphi, p_z, p_{\varphi}) \in (-a, b) \times {\mathcal{S}}^1 \times \mathbb{R}^2\, |\, \tilde H(z, \varphi, p_z, p_{\varphi})=h\,, \,\tilde J(z, \varphi, p_z, p_{\varphi})=c \right\} \nonumber \\
&
=\left\{ (z,  \varphi, p_z, p_{\varphi}) \in (-a, b) \times {\mathcal{S}}^1 \times \mathbb{R}^2\, \Big |\, 
\frac{n}{2} 
\left( \frac{p_z^2}{1+f'^2(z)} + \frac{p_{\varphi}^2}{f^2(z)} \right) +W(z)
=h\,, \,p_{\varphi}=c \right\}\,\nonumber \\
&=
\left\{ (z, p_z, p_{\varphi}) \in (-a, b) \times \mathbb{R}^2\, \Big |\, 
\frac{n}{2} 
\left( \frac{p_z^2}{1+f'^2(z)}  \right) +W_c(z)
=h\,, \,p_{\varphi}=c \right\} \times {\mathcal{S}}^1\,.
\label{level_2}
\end{align}
We will discuss some examples in Section \ref{sect:applications}.


\section{Lagrangian homographic relative equilibria}
 
\subsection{Some general existence criteria}
 
\bigskip
\noindent
The RE of the  dynamics on the ( homographic manifold 
correspond to RE of the full system with the particles form a regular $n$-ring at  situated on a plane perpendicular to the axis of rotation; we call these \underline{\textit{Lagrangian homographic RE}}. 
 In order to determined these it  is sufficient to determine their  base points in the $z$-direction. Indeed, once a such a base point $z_0$ is determined, the RE  solution is
\begin{equation}
z(t)=z_0\,, \quad p_z(t)=0\,, \quad \varphi(t)= \varphi_0+ \frac{nc}{f^2(z_0)}\,, \quad  p_{\varphi}(t)=c
\end{equation}
where we used the RE definition \eqref{def_RE},  the angular momentum conservation \eqref{total_ang_momentum_map} and the reconstruction equation \eqref{reconstruction}.
%

The base points may be found as  the critical points of the augmented potential $W_c(z)$ (see \cite{Ma92}). Since 
\begin{equation}
\sum \limits_{1\leq i<j\leq n} G\left(\frac{2k(j-i)\pi}{n}\, f(z)  \right) = \sum \limits_{k=1}^{n-1}(n-k) G\left(\frac{2k\pi}{n}\, f(z)  \right) 
\end{equation}
we can write 
\[
W_c(z)= \frac{n}{2} \frac{c^2}{f^2(z)} +\sum \limits_{1\leq i<j\leq n} G\left(\frac{2k(j-i)\pi}{n}\, f(z)  \right)=\frac{n}{2} \frac{c^2}{f^2(z)} +  \sum \limits_{k=1}^{n-1}(n-k) G\left(\frac{2k\pi}{n}\, f(z)  \right)
\]
and so we have
\begin{align}
W'_c(z)=
  f'(z)\left(-\frac{nc^2}{f^3(z)}  +  \sum \limits_{k=1}^{n-1} \frac{2(n-k)k\pi}{n}G'\left(\frac{2k\pi}{n}\, f(z)  \right) \right)\,.   
\label{v_prim}
\end{align}
We immediately deduce that if $\mathbb{M}$ admits a  geodesic circle $z=z_0$ (and so $f'(z_0)=0$), then $z_0$ is critical points of $W_c(z)$ and thus a base point for a RE.

\smallskip

\begin{definition}
The   potential \eqref{initial_potential_1}   is \underline{attractive}  (\underline{repulsive}) if the function $G$, is such that  $G'(x)> 0$  ($G'(x)<0$) for  all $x>0$. 
\label{def_attractive} 
\end{definition}

Taking into account the formula of $W'_c(z)$ given by \eqref{v_prim}, and that $f(z)>0$ for all $z$, the next propositions are immediate.

\begin{proposition}
In the generalized $n$-body problem  on surfaces of revolution, if the masses are equal and the potential is repulsive, then the only  Lagrangian homographic RE with the shape of a regular  $n$-gon are those with trajectories   on the   geodesic circles.
\end{proposition}

\begin{corollary}
In the generalized $n$-body problem  on surfaces of revolution, if the masses are equal,  the potential is  repulsive and the profile curve $f(z)$ has no critical points,   then there are no Lagrangian homographic RE.
\end{corollary}
\begin{corollary}
In the generalized $n$-body problem  on $\mathbb{S}^2$, if the masses are equal and  the potential is  repulsive,   then the only  Lagrangian homographic RE   are those with their trajectories on the Equator. \end{corollary}

\begin{proposition}\label{RE_n_body_attracative}
In the generalized $n$-body problem  on surfaces of revolution, if the masses are equal and  if the  potential is  attractive, then any  parallel circle (not necessarily geodesic) is a trajectory of  a Lagrangian homographic RE. 
\end{proposition}

\subsection{Bifurcations of geodesic Lagrangian homographic  RE and  curvature}

In this subsection we show that for attractive potentials,  a Lagrangian homographic  RE with its trajectory on a geodesic circle generically experiences a pitchfork bifurcation as the angular momentum vary. 
%

%





To start, recall  that in a two-degrees of freedom Hamiltonian  mechanical system with rotational symmetry, the stability modulo rotations of a RE  may be  determined the sign of second derivative  of the augmented potential at the base point. In our context, let $z_0$ be a base point for a RE. Then:
\begin{itemize}
\item if $W''_c(z_0) <0$ then the RE is unstable (and thus unstable in the full phase-space);

\item  if $W''_c(z_0) >0$ then the RE is stable within the dynamics on invariant homographic manifold only;

\item if $W''_c(z_0) =0$ then the RE stability is undecided and generically $z_0$ is the landmark of a bifurcation.
\end{itemize}
%

Consider a  Lagrangian homographic RE with its trajectory on the geodesic circle $z=z_0$.
Given the formula \eqref{v_prim} of $W'_c(z)$ and that $f'(z_0)=0$ we calculate
\begin{equation}
W''_c(z_0) =  f''(z_0)\left(-\frac{nc^2}{f^3(z_0)}  +  \sum \limits_{k=1}^{n-1} \frac{2(n-k)k\pi}{n}G'\left(\frac{2k\pi}{n}\, f(z_0)  \right) \right)
\label{second_w}
\end{equation}
On the other hand, recall that for a given a surface of revolution generated by a function $f(z)$ the Gaussian curvature is 
\begin{align}
K(z):= -\frac{f''(z)}{f(z)\left( 1+ f'^2(z) \right)^2}
\label{Curvature_K}
\end{align}
Thus the curvatures changes its sign as the second derivative of $f$ does. Considering the expressions of $W''_c(z_0)$ and $K(z)$ above, we  immediately obtain

\begin{proposition}
Consider  a  Lagrangian homographic  RE with its trajectory on the geodesic circle $z=z_0$. If the Gaussian curvature at $z_0$ is negative (i.e. $K(z_0)<0$ and so $f''(z_0)>0$) and $G$ is repulsive (and so $G'(x)<0$ for all $x$), then the RE is unstable.
\end{proposition}

For motions with $G$ attractive, we have:

%
\begin{proposition}[Geodesic Lagrangian homographic RE bifurcation criterion]
\label{pitchfork}
Consider  a  Lagrangian homographic RE with its trajectory on a non-zero curvature geodesic circle $z=z_0$,   and assume that $G$ is attractive. If
 %
%
%
%
%
\begin{equation}
 \sum \limits_{k=1}^{n-1} 
2(n-k)k
 \left[
3 G'\left(\frac{2k\pi}{n} \, f(z_0)\right) +2k\pi G''\left(\frac{2k\pi}{n}\, f(z_0)  \right)
  \right] \neq 0
 \label{condition} 
\end{equation}
then $(z_0,c_0)$ is the landmark of a pitchfork bifurcation.

\end{proposition}

\noindent 
Proof:  Let us denote $W_{c} (z) =  W(z, c)$. Standard bifurcation theory for one-degree of freedom Hamiltonians  (see, for instance, \cite{HH07}) guarantees the existence of a pitchfork bifurcation  at a point $(z_0, c_0)$ if 
\begin{equation}
\frac{\partial^2 W(z, c)}{\partial z^2}\Big|_{(z_0, c_0)} = \frac{\partial^3 W(z, c)}{\partial z^3}\Big|_{(z_0, c_0)}  =\frac{\partial^2 W(z, c)}{\partial c \partial z} \Big|_{(z_0, c_0)} =0\,,
\label{zero_cond}
\end{equation}
and 
\begin{equation}
 \frac{\partial^3 W(z, c)}{\partial c \partial z^2}\Big|_{(z_0, c_0)} \neq 0\,,  \quad \quad \frac{\partial^4 W(z, c)}{\partial z^4}\Big|_{(z_0, c_0)} \neq 0\,.
 \label{W_4}
\end{equation}
For reader's convenience we re-write formula \eqref{v_prim} of $\displaystyle{W'_c(z) \equiv \frac{\partial W(z,c)}{\partial z}}$ below:
\begin{equation}
\frac{\partial W(z,c)}{\partial z} =  f'(z)\left(-\frac{nc^2}{f^3(z)}  +  \sum \limits_{k=1}^{n-1} \frac{2(n-k)k\pi}{n}G'\left(\frac{2k\pi}{n}\, f(z)  \right) \right)\,.   
\label{partial_prim}
\end{equation}
%
%
%
%
%
%
Let $c_0>0$ be the positive root of the parenthesis on the right hand side of the \eqref{partial_prim} above, and so
\begin{equation}
c_0^2 = \frac{f^3(z_0)}{n} \sum \limits_{k=1}^{n-1} \frac{2(n-k)k\pi}{n}G'\left(\frac{2k\pi}{n}\, f(z_0)  \right)\,.  
\label{c_zero_here}
\end{equation}
Taking into account that $z_0$ is such that $f'(z_0)=0$ and that $(z_0,c_0)$ cancels the right hand side parentheses  of \eqref{partial_prim}, one verifies that  conditions \eqref{zero_cond} are fulfilled and that

\begin{equation}
\frac{\partial^3 W(z, c)}{\partial c \partial z^2}\Big|_{(z_0, c_0)}  =-\frac{2n c_0 f''(z_0)}{f^3(z_0)} \neq 0
\end{equation}
and 
\begin{equation}
 \frac{\partial^4 W(z, c)}{\partial z^4}\Big|_{(z_0, c_0)} = 3 \left(f''(z_0)\right)^2 \left[ \frac{3n c_0^2}{f^4(z_0)}  +   \sum \limits_{k=1}^{n-1} \frac{4(n-k)k^2\pi^2}{n^2} G''\left(\frac{2k\pi}{n}\, f(z_0)  \right)  \right] \neq 0\,.
\end{equation}
Using \eqref{c_zero_here}, we substitute $c_0^2$ into the above and get  
\begin{equation}
 \frac{\partial^4 W(z, c)}{\partial z^4}\Big|_{(z_0, c_0)} = 3 \left(f''(z_0)\right)^2 \frac{\pi}{n^2} \left( \sum \limits_{k=1}^{n-1} 
2(n-k)k
 \left[
3 G'\left(\frac{2k\pi}{n} \, f(z_0)\right) +2k\pi G''\left(\frac{2k\pi}{n}\, f(z_0)  \right)
  \right] \right)    \,.
\end{equation}
Since the curvature at $z=z_0$ is non-zero  (and so $f''(z_0) \neq 0$) and using the condition  \eqref{condition},  the conclusion follows. $\square$
%
%
 %
\begin{remark}
\label{typical}
It seems likely  the inequality  \eqref{condition}  is fulfilled by most attractive potentials and for most values of $n$. In this sense, this condition is typical for most $n$-body problems on a surface or revolution.  A conjecture stated for the specific case of the 3-d gravitational $n$-body problem $\mathbb{S}^2$ will be stated later (Conjecture \ref{conjecture}).
\end{remark}

\begin{remark}
\label{inner_branch}
If in the conditions \eqref{W_4}  we have $\frac{\partial^4 W(z, c)}{\partial z^4}\Big|_{(z_0, c_0)} >0$,  then the pitchfork bifurcation is subcritical, that is the inner branch of RE (in between the outer branches, when they exist) is unstable. Note that $\frac{\partial^4 W(z, c)}{\partial z^4}\Big|_{(z_0, c_0)} >0$ is equivalent to 
\begin{equation}
 \sum \limits_{k=1}^{n-1} 
2(n-k)k
 \left[
3 G'\left(\frac{2k\pi}{n} \, f(z_0)\right) +2k\pi G''\left(\frac{2k\pi}{n}\, f(z_0)  \right)
  \right] > 0\,.
 \label{subcritical_pitch} 
\end{equation}
\end{remark}

\begin{remark}\label{z_remark}
\label{outer_branches}
The trajectories of the  outer branches RE emanating at the  pitchfork bifurcation point are on  parallel circles $z=\pm z(c)$  which are found by solving  the right hand side  parenthesis of   \eqref{partial_prim} equal to zero for $|c| \neq c_0.$ Specifically, 
$(\pm z(c), c)$  are points on the curve 
\begin{equation}
F(z,c):=-\frac{nc^2}{f^3(z)}  +  \sum \limits_{k=1}^{n-1} \frac{2(n-k)k\pi}{n}G'\left(\frac{2k\pi}{n}\, f(z)  \right) =0\,.  
\label{z_biff_RE}
\end{equation}
where  $|c|\neq c_0.$
From the conditions assuring the existence of the pitchfork bifurcation at $|c|=c_0$, we know that the equations above has precisely two roots  symmetrically disposed with respect to, and in a neighbourhood of, $z_0$.
\end{remark}


Taking into account the relation \eqref{Curvature_K}   between $f''(z)$ and the curvature $K(z)$, and given the equation \eqref{second_w} for $W''_c(z_0)$, we also deduce:
\begin{proposition}[Unstable geodesic Lagrangian homographic RE]
\label{unstable_criterion}
Consider  a Lagrangian homographic RE with its trajectory on geodesic circle $z=z_0$ and assume that $G$ is attractive. Let  $c_0>0$  be the momentum value which cancels the  parenthesis in the right hand side of \eqref{second_w}. 

\begin{enumerate}

\item
If the curvature at $z_0$ is positive (i.e. $K(z_0)>0$ and so $f''(z_0)<0$), then
\begin{enumerate}
\item for $|c|<c_0$ the RE is unstable;

\item for $|c|\geq c_0$ the stability of the RE is undecided. More precisely, for $|c|>c_0$ the RE is stable within the homographic invariant manifold only.

\end{enumerate}
\item
If the curvature at $z_0$ is negative (i.e. $K(z_0)<0$ and so $f''(z_0)>0$), then
\begin{enumerate}
\item for $|c|\leq c_0$ the stability of the RE is undecided. More precisely, for $|c|>c_0$ the RE is stable within the homographic invariant manifold only. The momentum $|c|=c_0$ generically marks a pitchfork bifurcation;

\item for $|c|> c_0$ the RE is unstable.

\end{enumerate}
\end{enumerate}

\end{proposition}

\section{Some examples}
\label{sect:applications}


\subsection{Homographic dynamics for the ``quasi-harmonic" interaction}

We discuss  homographic motion with a ``quasi-harmonic"  binary interaction given by $\displaystyle{G(x) = x^2/2}$, $G:[0, \infty) \to \mathbb{R}$. Note that $G'(x)=x>0$ for all $x>0$ so $G$ is attractive.
The amended potential is
\begin{align}
W_c(z) = \frac{nc^2}{2 f^2(z)} + \sum\limits_{1\leq i<j\leq n} \frac{1}{2}  \left(\frac{2\pi(j-i)}{n} f(z)  \right)^2\,.
\end{align}
After some elementary algebra, given that
\begin{align}
\sum\limits_{1\leq i<j\leq n}  (j-i)^2=\sum\limits_{k=1}^{n}  \frac{(k-1)k(2k-1)}{6}
=\frac{n^2(n+1)^2(n+2)}{24}
\end{align}
we have
\begin{align}
W_c(z) = \frac{nc^2}{2f^2(z)} + \frac{\pi^2(n+1)^2(n+2)}{12}f^2(z)\,,
\label{amended_quasi-harmonic}
\end{align}
and so
\begin{align}
W'_c(z) = f'(z)\left( -\frac{nc^2}{f^3(z)} + \frac{\pi^2(n+1)^2(n+2)}{6}f(z)\right)\,.
\end{align}
Assume that $z=z_0$ is a  geodesic circle and so $f'(z_0)=0$. The pitchfork bifurcation   in Proposition \eqref{pitchfork} is retrieved   at 
\begin{equation}
c_0=c_0(n):=\pi (n+1)\sqrt{\frac{n+2}{6n}f^4(z_0)}
\label{c_zero_general}
\end{equation}
and we distinguish:
\begin{enumerate}
\item for $c=0$ there is a unique equilibrium on the parallel geodesic circle $z=z_0$;

\item for every $|c|$ such that
\begin{enumerate}

\item $|c| < c_0(n)$ if $f''(z_0)<0$ (i.e. $z=z_0$ has positive curvature)

\item $|c| > c_0(n)$ if $f''(z_0)>0$ (i.e. $z=z_0$ has negative curvature)
 
\end{enumerate}

\noindent
there are three RE, one with its trajectory  on $z=z_0$ and the other two with their trajectories   symmetrically disposed (see  Remark \ref{z_remark}) on parallel circles at $\pm z(c)$, where the later $z$ values  solve 
\begin{align}
f^3(z) = \frac{6nc^2}{\pi^2(n+1)^2(n+2)}\,.
\label{roots_z}
\end{align}
in a neighbourhood of $z=z_0;$
%

\item for $|c| = c_0(n)$ 
%
%
%
there is a  RE and a pitchfork bifurcation;

\item for 

\begin{enumerate}

\item $|c| > c_0(n)$ if $f''(z_0)<0$ (i.e. $z=z_0$ has positive curvature)

\item $|c| < c_0(n)$ if $f''(z_0)>0$ (i.e. $z=z_0$ has negative curvature)

\end{enumerate}
there is a unique RE with its trajectory on $z=z_0.$ 

\end{enumerate}
\begin{figure}[h!]
\centering
           \includegraphics[angle=0,scale=0.5] {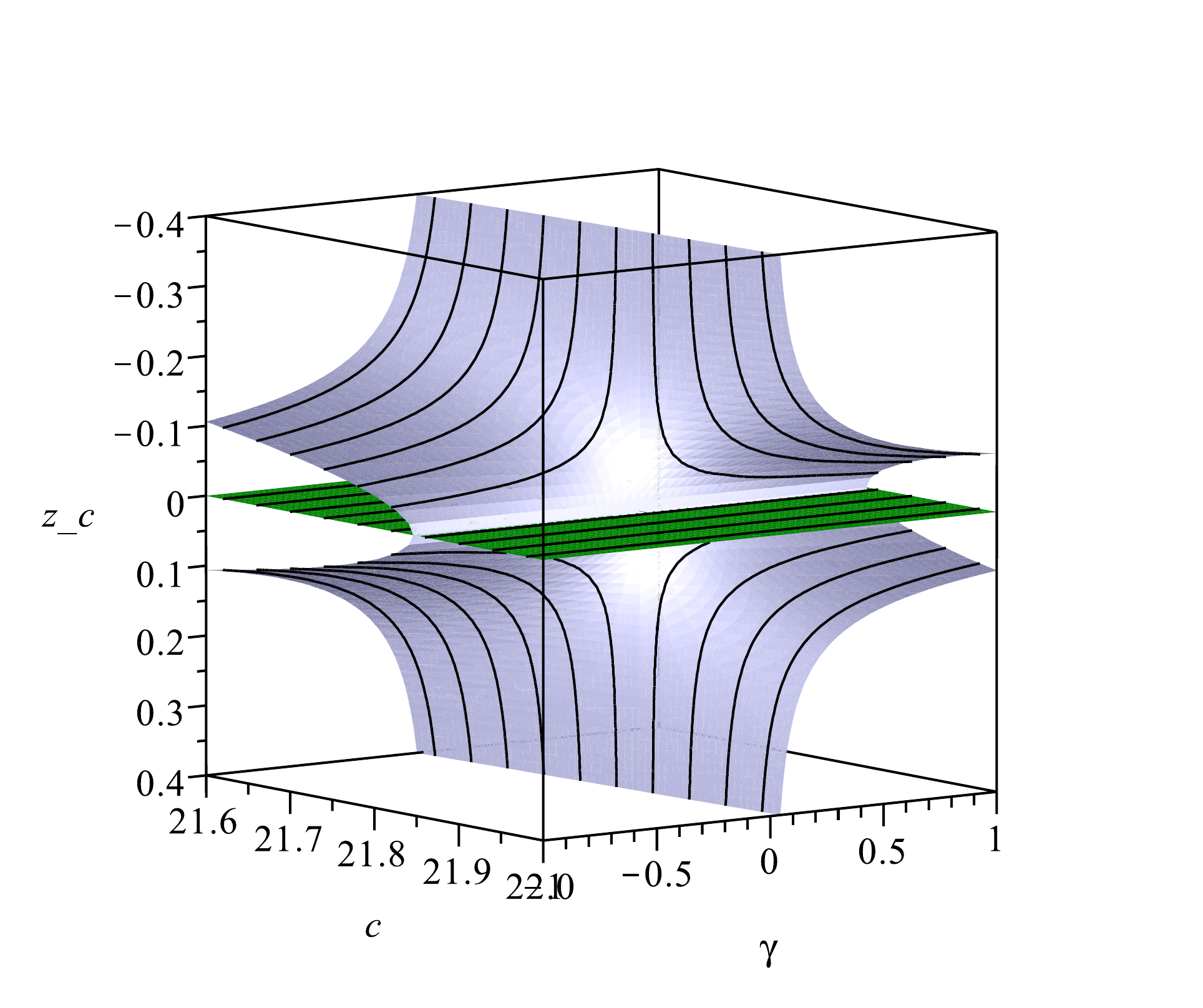}
           \caption{The coordinates $z= z(c, \gamma)$ (the indigo surface) of the pitchfork bifurcation of RE and $z=0$ (the green plane)  in the case of the quasi-harmonic potential for family of surfaces $\mathbb{M}$ generated by $f(z)=\sqrt{1+\gamma z^2},$ $z\in (-1,1),$ $\gamma \in [-1,1].$ For $\gamma = -1$   the surface $\mathbb{M}$ is a sphere and so its  curvature constant and positive. For $\gamma \in (-1, 0)$ the curvature is positive. At $\gamma=0$ the sign of the curvature changes and for $\gamma>0$  the surface $\mathbb{M}$ becomes a hyperboloid with one sheet. No matter the curvature sign, the inner RE  (that is the RE  in between branches when the branches exit) is unstable.  In this plot $n=15$. } 
         \label{RE_quasi-harmonic_3d}
\end{figure}

\subsubsection{Quasi-harmonic homographic motion on $\mathbb{S}^2$}

\begin{figure}[h!]
\centering
           \includegraphics[angle=0,scale=0.45] {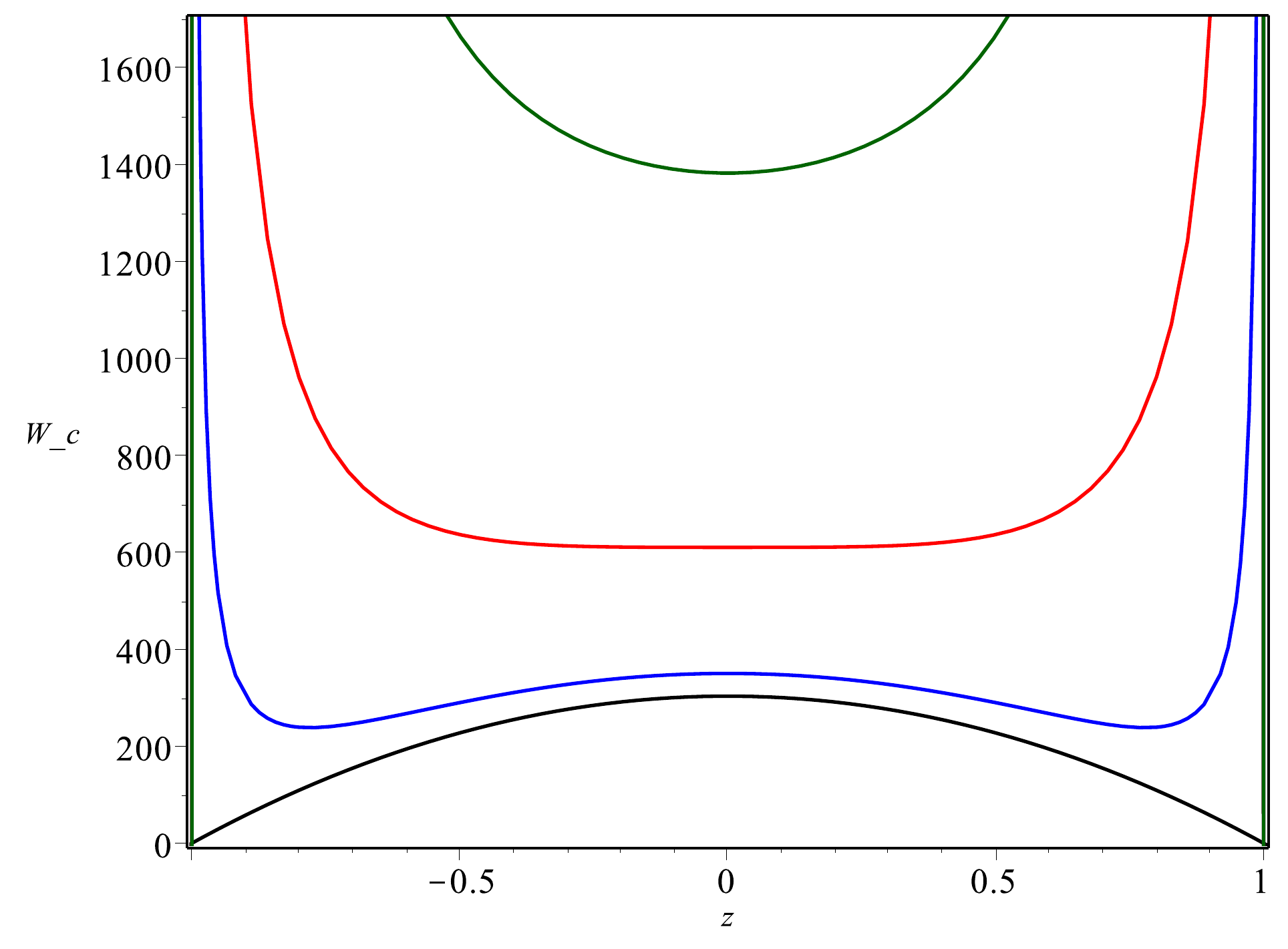}
           \caption{The amended potential  $W_c(z)$ of the homographic dynamics on $\mathbb{S}^2$. We have $z\in (-1,1)$, with $z=1$ marking the North Pole. The black, blue, red and green curves correspond to momenta $c=0$, $|c|\in (0, c_0(n))$, $|c|=c_0(n)$ and $|c|>c_0(n)$, respectively. The Hill regions of motions in the reduced phase-space $\{z, p_z\}$ are  defined by $\{z\,|\, W_c(z)\leq h\}$.    Various energy levels $h$ are represented by dotted horizontal lines.
           For $|c|<c_0(n),$ the Equatorial RE $z=0$ is unstable (in concordance to Proposition \ref{unstable_criterion}).} 
         \label{Figure_quasi-harmonic}
\end{figure}

\bigskip
We now focus  on  motion on the unit sphere $\mathbb{S}^2$.  The  generatrix of $\mathbb{S}^2$ is $f(z)=\sqrt{1-z^2}$ and for  $c\neq 0$, the reduced dynamics is given by 
\begin{align}
&H: (-1,1) \times {\mathbb S}^1 \times \mathbb{R} \times \mathbb{R} \to \mathbb{R}
\nonumber \\
&H(z,\varphi, p_z, p_{\varphi}) =  
\frac{n}{2}(1-z^2)p_z^2 + \frac{nc^2}{2(1-z^2)} + \frac{\pi(n^2-1)}{3n}(1-z^2)
\label{Ham_quasi-harmonic_n} 
\end{align}
whereas for $c=0$
\begin{align}
&H: [-1,1] \times {\mathbb S}^1 \times \mathbb{R} \times \mathbb{R} \to \mathbb{R}
\nonumber \\
&H(z,\varphi, p_z, p_{\varphi}) =
\frac{n}{2} (1-z^2)p_z^2 +  \frac{\pi(n^2-1)}{3n}(1-z^2)\,.
\label{Ham_zero_quasi-harmonic_n} 
\end{align}
%
%
%
%
%
%
When $c=0,$ the equations of motions are
\begin{align}
\dot z= 2(1-z^2)p_z\,,  \quad \quad \dot p_z= -
2z \left( \frac{n}{2}p_z^2 +  \frac{\pi(n^2-1) }{3n} \right)
\end{align}
and we observe that the boundaries $z=\pm1$ become fictitious invariant manifolds.
On the Equator (i.e., at $z=0$) the momentum bifurcation value \eqref{c_zero_general} becomes
\begin{equation}
c_0(n)= \pi (n+1)\sqrt{\frac{n+2}{6n}}
\end{equation}
whereas the values of the RE coordinates  as given by \eqref{roots_z} are
\begin{align}
\pm z(c):=\pm\sqrt{1- \frac{|c|}{\pi(n+1)}\sqrt{  \frac{6n}{n+2}  }  }
\label{z_0(n; c)}\,.
\end{align}
\bigskip
The topology of the phase space is easily deduced by applying the definition  \eqref{level_2} to the present context and the analysis of Figure \ref{Figure_quasi-harmonic}. Thus the  phase space of the homographic invariant manifold is foliated by:
\begin{enumerate}
\item for  $c=0$
\[
I_{h\,,0} = \begin{cases}
 \text{the void set} \,,\,\,\, & \text{if} \,\,h < 0\,, \\
 \text{two lines} \,,\,\,\, & \text{if} \,\,h = 0\,, \\
    \text{two identical strips, (each} \simeq
  \{
  \text{line} \times \text{closed interval}\})\,,  
 & \text{if} \,\,h\in (0\,, W_0(0))\,,\\
 \text{two identical strips glued together}\,,
& \text{if} \,h= W_0(0)\,, \\
    \text{a strip}\,,    & \text{if} \,\,h>W_0(0)\,. \\
\end{cases}
\]
\item for $|c| \in (0\,, {c_0(n)})$ 
\[
I_{h\,,c} = \begin{cases}
    \text{the void set} \,,\,\,\, & \text{if} \,\,h < W_c(z(c)) \,,\\
     \text{a circle} \,,\,\,\, & \text{if} \,\,h = W_c(z(c)) \,,\\
    \text{two disjoint 2-tori (each $\simeq S^1 \times S^1$)}\,, \,,\,\,\, & \text{if} \,\,h \in (W_c(z(c)), W_{c} (0))\,, \\
    \text{two 2-tori tangent to each other along a big circle}\,,  & \text{if} \,\,h= W_{c} (0) \,, \\
    \text{one 2-torus}\,,   & \text{if} \,\,h> W_{c} (0)\,.
\end{cases}
\]
\item for $|c| \geq  {c_0(n)}$ 
\[
I_{h\,,c} = \begin{cases}
    \text{the void set} \,,\,\,\, & \text{if} \,\,h < W_c(z(c))\,,\\
    \text{a circle}\,,  & \text{if} \,\,h= W_c(z(c))\,, \\
    \text{a 2-torus}\,,   & \text{if} \,\,h> W_c(z(c)).
\end{cases}
\]
\end{enumerate}

\begin{figure}[h!]
\centering
      \subfigure[The phase curves for  $c=0.$]
      {
           \includegraphics[angle=0,scale=0.26] {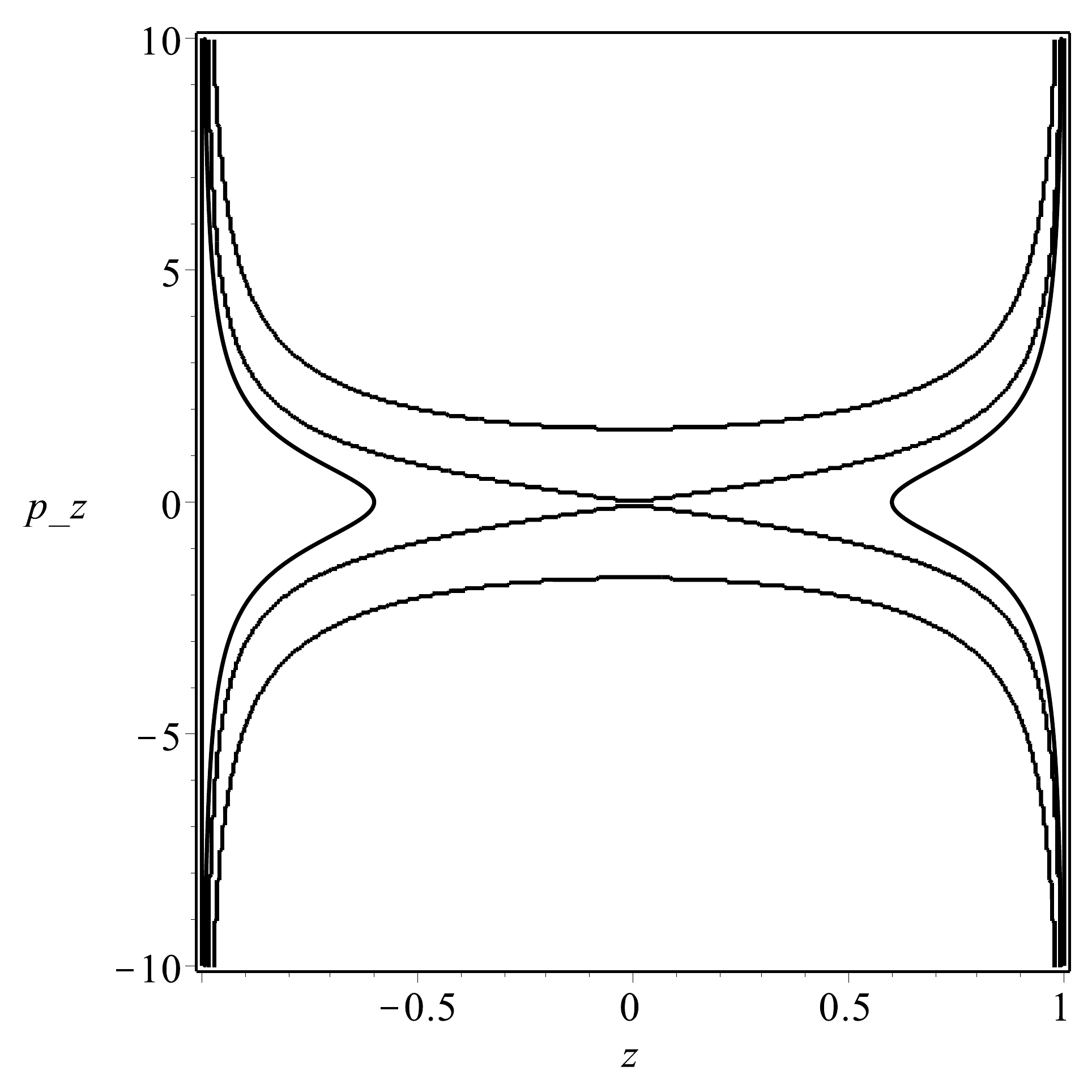}
           } \quad 
      \subfigure [The phase curves for  $|c|\in (0, c_0).$]
       {
            \includegraphics[angle=0,scale=0.26] {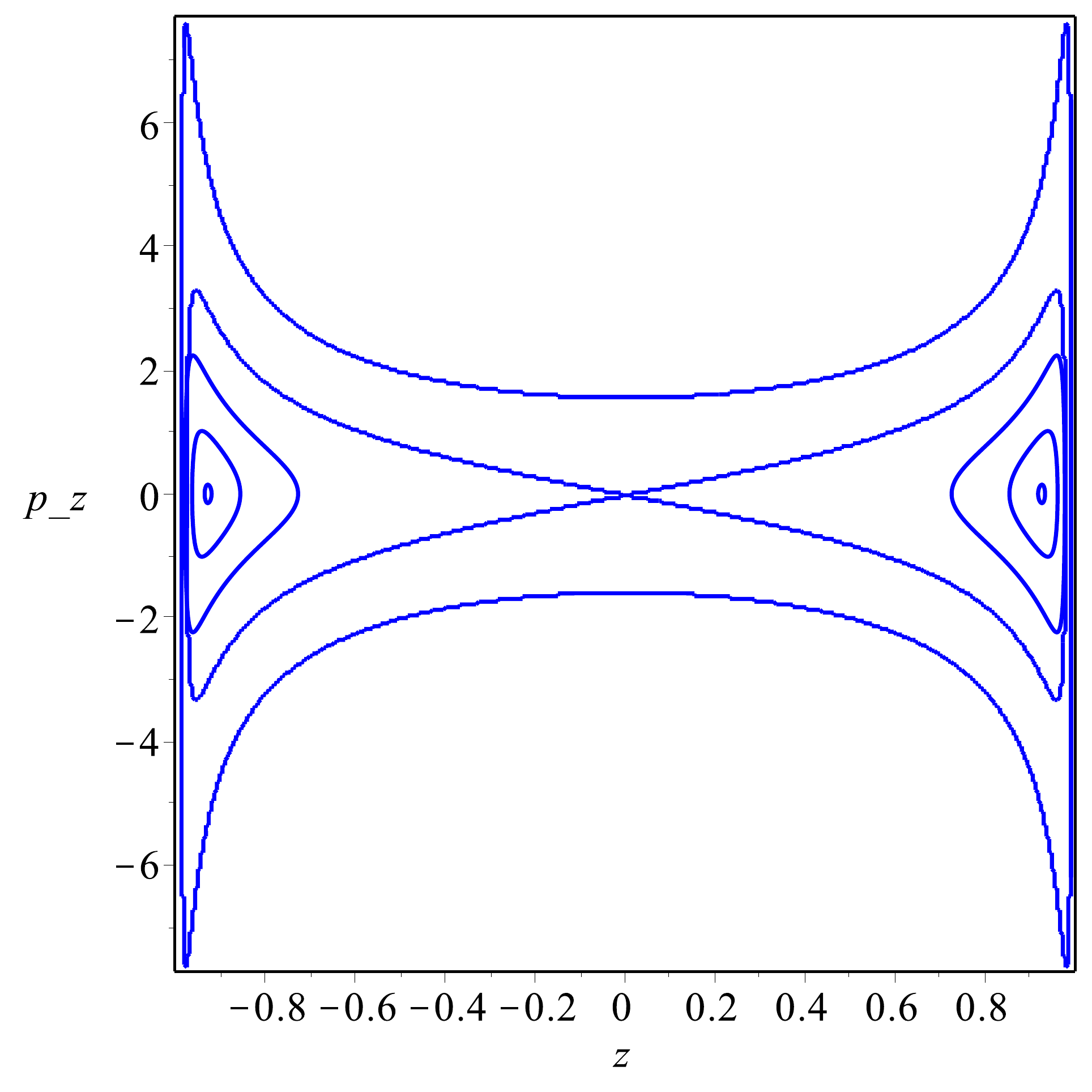}
         } \quad 
         \subfigure [The phase curves for  $|c| \geq c_0.$]
       {
            \includegraphics[angle=0,scale=0.26] {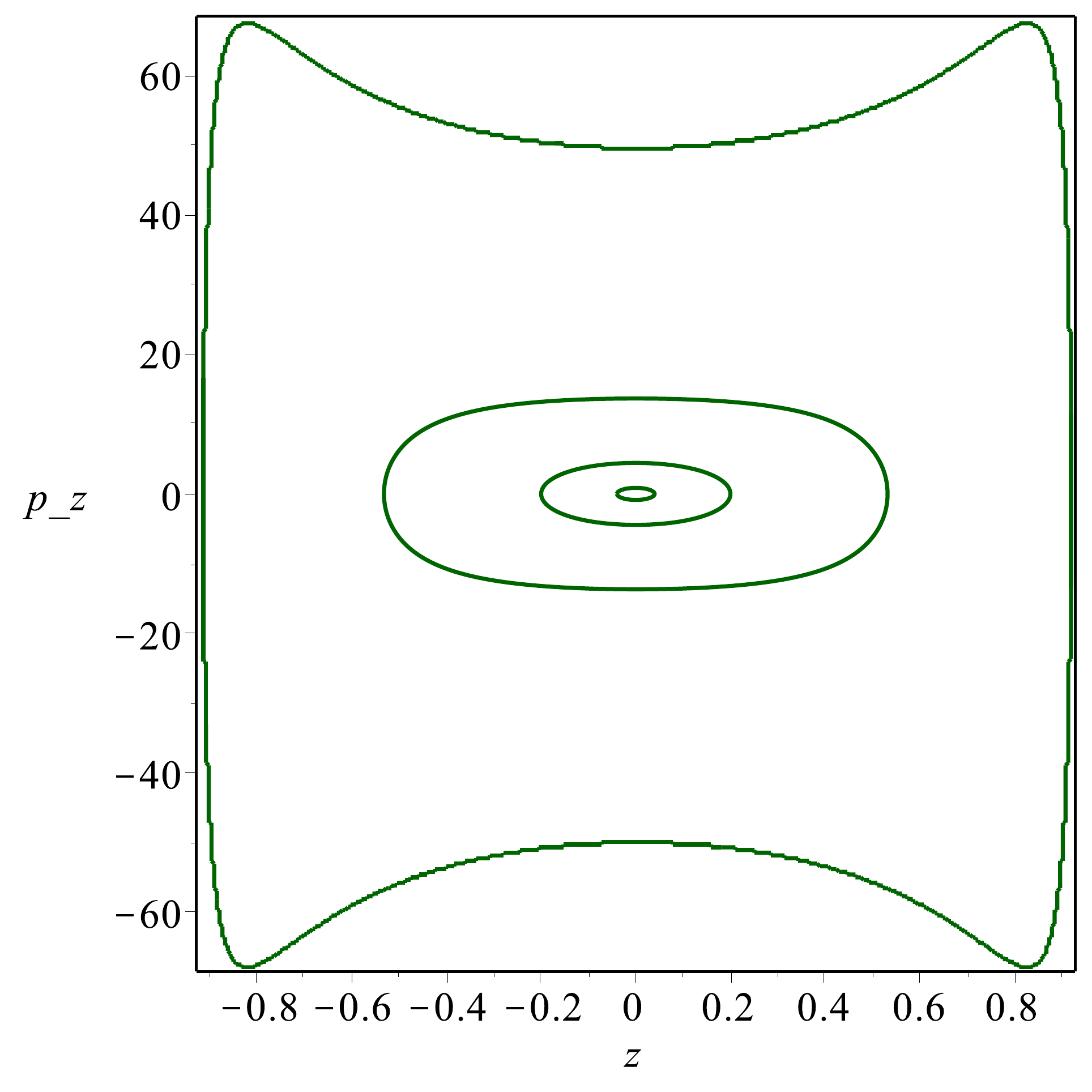}
         } 
         \caption{The  phase plane of the homographic invariant manifold for particles in quasi-harmonic interaction constrained to move on $\mathbb{S}^2$.}
         \label{Figure_flow_quasi-harmonic}
\end{figure}

\bigskip


\begin{proposition}\label{prop-quasi-harmonic} Consider  $n$ particles constrained to move on $\mathbb{S}^2$ and with pairwise  quasi-harmonic interaction. As  the energy $h$ and the momentum $c$ vary, the dynamics on the homographic  Lagrangian invariant manifold consists of:

\begin{enumerate}

\item for  $c=0$ 

\begin{enumerate}

\item for  $h < 0$ there is no dynamics;

\item for $h=0$ we have fictitious motion at the Poles $z=\pm1;$ these points, while  not singularities for the equations of motion, correspond to non-physical configurations (all points clustered in one point);

\item  for  $h \in (0,W_0(0) )$  we have homothetic motion (each particle moving on a meridian) consisting in orbits  
which are ejected from the North pole ($z=1$), reach a minimum value $z_{min}= z_{min}(h)$ and return to  the North pole; also, by  symmetry with respect to the Equator, we have  identical dynamics  in the Southern hemisphere; 

\item  for  $h =W_0(0)$  we have homothetic motion (each particle moving on a meridian) consisting in orbits  
which are ejected from the North pole ($z=1$) and tend to the Equator; by  symmetry we have  identical dynamics  in the Southern hemisphere;  also, there is an equilibrium on the Equator. 

\item  for  $h >W_0(0)$  we have homothetic motion (each particle moving on a meridian) consisting in orbits  
which are ejected from the North pole ($z=1$) and tend to the South Pole;  we have  identical dynamics  in the Southern hemisphere; 

\end{enumerate}

\item for  $|c|\in(0\,, c_0(n))$,  let $z(c)$ be defined as in Remark \ref{z_remark}. Then

\begin{enumerate}

\item for $h = W_c(z(c))$ there is no dynamics;

\item for $h = W_c(z(c))$ there are two RE with their trajectories located at $z=\pm z(c);$  

\item   for $h \in \left(W_c(z(c)), W_{c} (0) \right)$, in the Northern  hemisphere the orbits are contained in  
a parallel annular sector  $z \in [z_{min}\,, z_{max}]$, where $z_{min}$ and $z_{min}$ depend on the energy level. These orbits may be periodic or quasi-periodic, filling densely the  annular sector.   By  symmetry, we have  identical dynamics  in the Southern hemisphere;

\item  for  $h =  W_{c} (0)$  in the Northern  hemisphere, the orbits reach an extreme value of $z$ and spiral asymptotically towards the RE on the Equator;  identical dynamics are in the Southern hemisphere.


\item   for $h >  W_{c} (0)$  the particles visit, in a symmetric way, both hemispheres crossing the Equator, with $z$ in between  maximal and minimum values which depend on the energy level;

\end{enumerate}

\item for $|c| \geq c_0$

\begin{enumerate}
\item for $h< W_{c}(0)$ there is no dynamics;

\item  for   $h= W_{c}(0)$    we have a RE with its trajectory on the Equator;

\item  for   $h> W_{c}(0)$  we have orbits similar to those in the case 2.(e).

\end{enumerate}

\end{enumerate}

\end{proposition}

\noindent Proof: This follows from a a direct analysis of the one-degree of freedom Hamiltonian  \eqref{Ham_quasi-harmonic_n}. See also the sketch of the amended potential in in Figure \ref{Figure_quasi-harmonic} and the phase curves in Figure \ref{Figure_flow_quasi-harmonic}.
 $\square$

\subsubsection{Quasi-harmonic homographic motion  on a peanut-like surface}

{ \begin{wrapfigure}{r}{6cm}
\label{wrap-fig:1}
\includegraphics[width=5cm, scale=5]{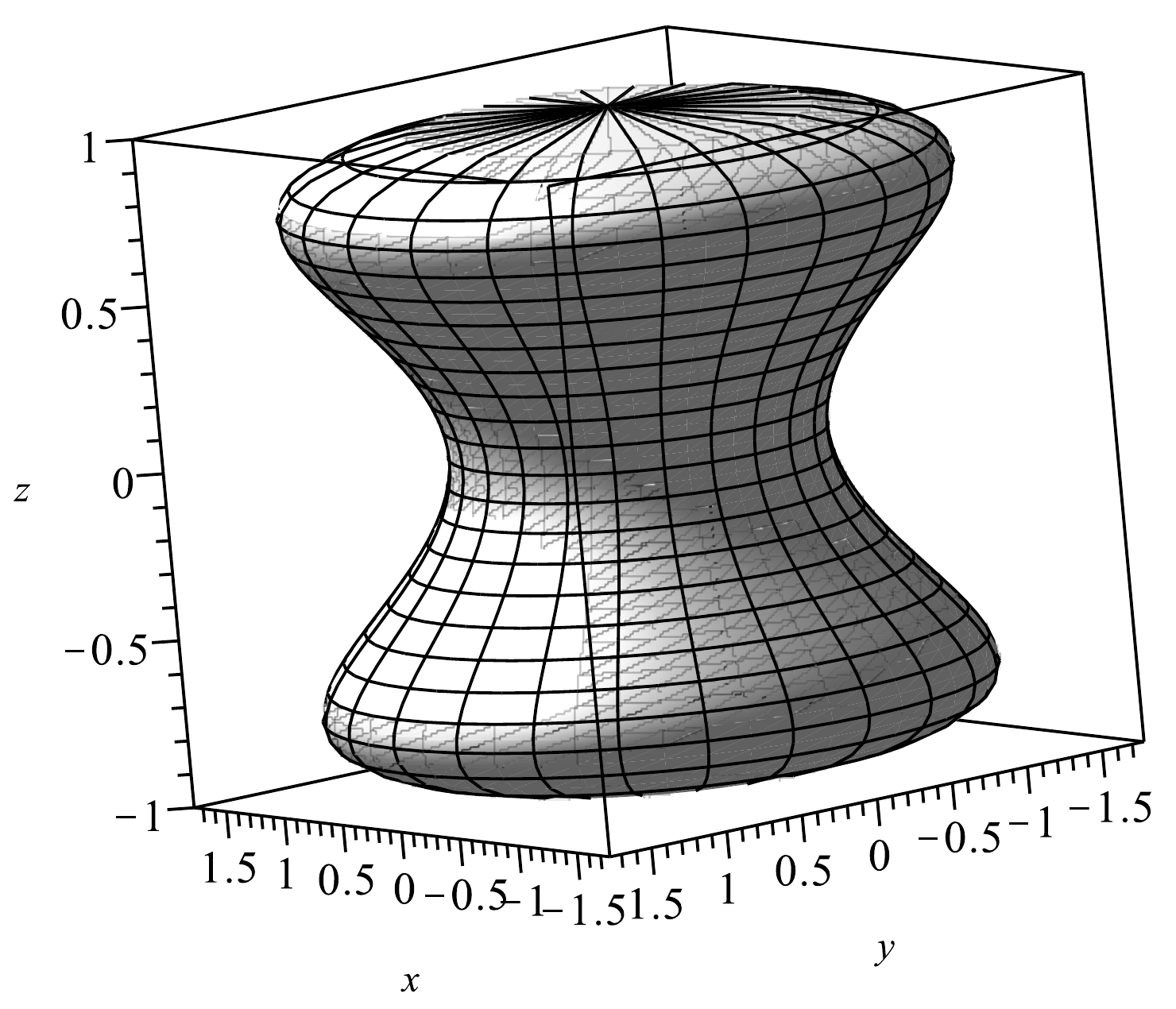} 
\end{wrapfigure} 
  {{$\,$\\  %
We consider now homographic dynamics on a surface $\mathbb{M}$ generated by
\begin{equation}
f(z) = (1+\gamma z^2)\sqrt{1-z^2}\,,\,\, \quad z\in (-1,1)\,,\,\, \,\,\, \gamma>\frac{1}{2}\,.
\nonumber
\end{equation}
The sign of curvature $K=K(z, \gamma)$ of $\mathbb{M}$ is that of 
\begin{equation*}
P(z, \gamma) :=  z\left(6 \gamma z^4 -9\gamma z^2 +2(\gamma -1) \right)
\end{equation*}
and so  the curvature $K(z,\gamma)$ cancels at
\begin{equation}
z=0 \quad \,\, \text{and} \quad z=\pm z_\gamma =: \pm \sqrt{  \frac{9\gamma - \sqrt{33 \gamma^2 +24 \gamma}}{12 \gamma}}
\label{z_gamma}
\end{equation}
}}
}
Further,
\begin{equation}
\begin{cases}
      K(z,\gamma)>0& \text{if}\,\,\, |z|< z_\gamma\,, \\
       K(z,\gamma)<0& \text{if}\, \,\,|z|> z_\gamma \,.
\end{cases}
\end{equation}
%
%
%
%
The amended potential is
\begin{align}
W_c(z) = \frac{nc^2}{2(1+\gamma z^2)^2(1-z^2)} + \frac{\pi^2(n+1)^2(n+2)}{12}(1+\gamma z^2)(1-z^2)\,,
\label{amended_peanut}
\end{align}
and, by formula \eqref{c_zero_general}, the momentum bifurcation values are
\begin{align}
& c_0= \pi (n+1) \sqrt{\frac{n+2}{6n}f^3(0)} =\pi (n+1) \sqrt{\frac{n+2}{6n}}\quad \,\,\,\,\,\text{for the geodesic circle}\,\,\,z=0 \\ 
& c_{\gamma}= \pi (n+1) \sqrt{\frac{n+2}{6n}f^3( z_\gamma)} \quad \,\,\,\,\,\text{for the geodesic circles}\,\,\,z=\pm z_\gamma\,.
\end{align}
Using elementary calculus it can be proven that $c_0<c_\gamma$ for $\gamma>1/2.$
\begin{figure}[h!]
\centering
           \includegraphics[angle=0,scale=0.55] {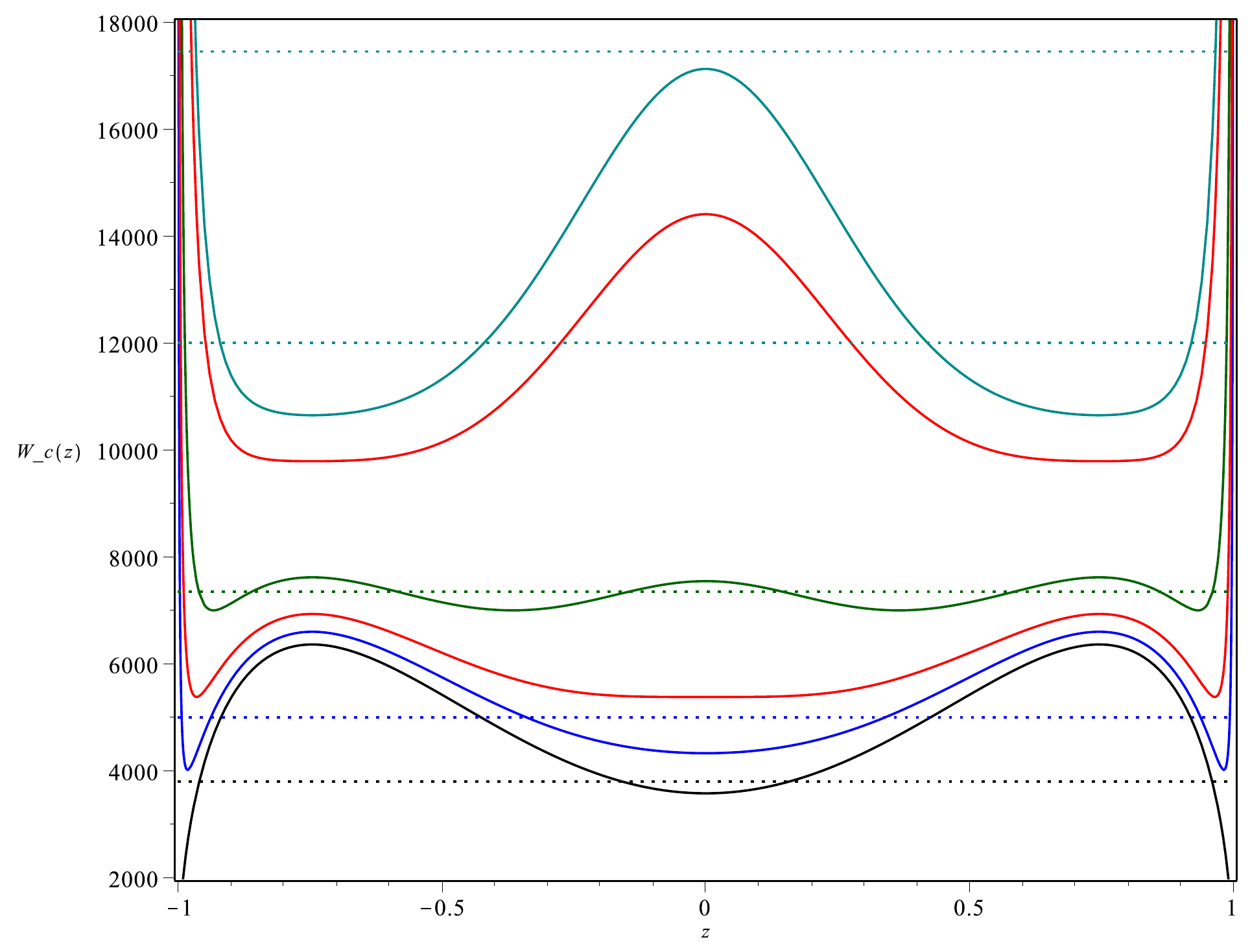}
           \caption{Plot of the amended potential \eqref{amended_peanut}. The amended potential  $W_c(z)$ of the homographic dynamics on a sphere. The  curves correspond to momenta $c=0$ (black), $|c|\in (0, c_0)$ (blue), $|c|=c_0$ (red), $|c|\in (c_0, c_\gamma)$  (light green) $|c|>c_\gamma$ (red), and $|c|> c_{\gamma}$ (dark green). The Hill regions of motions in the reduced phase-space $\{z, p_z\}$  defined by $\{z\,|\, W_c(z)\leq h\}$  for various energy levels $h$ (dotted horizontal lines).
           Note that  for $|c|>c_0,$ the Equatorial RE $z=0$ is unstable, and for $|c| \in (0, c_{\gamma})$ the outer RE are also unstable (in concordance to Proposition \ref{pitchfork}).} 
         \label{W_c_peanut}
\end{figure}
Given the sketch of the amended potential  - see Figure \ref{W_c_peanut} -  one may fully describe  the topology of the phase-space and the dynamics  by performing the analysis in  an analogous manner as for motion on $\mathbb{S}^2$. Instead, we choose to point out some of the main  features. Let us call the regions $z>0$ and $z<0$ the Northern and Southern hemisphere, respectively, and the geodesic parallel $z=0$ the Equator.  Note that the trajectories are symmetric with respect to the Equator. For appropriate levels of energy $h$ we have

\begin{enumerate}

\item for $c=0$ (in Figure \ref{W_c_peanut}, $W_c(z)$  corresponds to the black curve) 

\begin{itemize} 

\item there are homothetic trajectories connecting from the North to the South Pole, trajectories connecting a pole to a maximum distance,  and trajectories  going up and down in a symmetric manner with respect to the Equator (in Figure \ref{W_c_peanut}, regions bounded on the top by  the black dotted curve, on the bottom the black curve and on the sides by $z\pm 1$); 

\item there are three equilibria: two unstable, one in each hemisphere with their coordinates symmetrically disposed with respect to the Equator, and a third one on the Equator (in Figure \ref{W_c_peanut}, the critical points on the black curve);

\end{itemize}

\item for $c\neq 0$ (in Figure \ref{W_c_peanut}, $W_c(z)$  corresponds to the blue, red, light green, and dark green curves)

\begin{itemize}

\item  at each energy level there are three RE  with trajectories on each of the geodesic parallel circles, two with trajectories on $z=z_{\gamma}$ and one on $z=0$; they undergo three distinct pitchfork bifurcations as prescribed by Proposition \ref{pitchfork}

\item   the RE with trajectories on $z=z_{\gamma}$ (i.e., on geodesic circles with positive curvature) are unstable at low momenta, whereas the  Equatorial RE is unstable at high momenta (i.e., on geodesic circles with negative curvature); 

\item due to symmetry, two of the persistent RE appear always in pairs, one of each hemisphere, symmetrically disposed with respect to the Equator; 

\item there are momenta $c\neq 0$ for which, as $h$ vary, we have seven distinct RE (in Figure \ref{W_c_peanut}, the critical points on the light green curve);

\item there are levels of energy for which there are four distinct Hill regions, two on each hemisphere (in Figure \ref{W_c_peanut}, regions bounded on the top by  the green dotted curve and  at the bottom by the light green curve). In this case, in the phase space we have 4 distinct tori.

\end{itemize}

\end{enumerate}

\subsection{Gravitational homographic motion on $\mathbb{S}^2$ and $\mathbb{H}_{\text{one}}^2$}

In this Subsection we consider  homographic  for motions on $\mathbb{S}^2$ and  $\mathbb{H}^2_{\text{one}}$ for 3-d gravitational interactions. Recall that that 3-d gravitational potentials  are defined as  spherical-symmetric solutions of the Laplace equation 
on $\mathbb{S}^3$ ($\mathbb{H}^3$) and then restricted to the submanifolds $\mathbb{S}^2$ ($\mathbb{H}^2_{\text{one}}$). (For more on gravitational potentials on surfaces of constant curvature, see  \cite{Diacu2011, Vo05}.) For any two points $\q_i$  and $ \q_j$, the binary potential is
\begin{equation}
V(\q_i, \q_j)=- m_i m_j \cot \left(d(\q_i, \q_j) \right) \,\,\,  \text{on}  \,\,\,  \mathbb{S}^2 
\quad \quad \text{and}\quad \quad 
V(\q_i, \q_j)=- m_i m_j \coth \left(d(\q_i, \q_j) \right)   \,\,\,  \text{on}  \,\,\,  \mathbb{H}^2_{\text{one}}
\label{grav_pot_s}
\end{equation}
%

    

    Since $G(x)=-\cot x$, $G'(x) = 1+ \cot^2 x>0$ and so $G$ is attractive (in agreement with our Definition \ref{def_attractive}). By Proposition \ref{RE_n_body_attracative},  we have  that the 3-d gravitational $n$-body problem on $\mathbb{S}^2$ ($\mathbb{H}^2_{\text{one}}$)   admits Lagrangian homographic RE with Equatorial  trajectories. However, since configurations with points diametrically opposite on the Equator  are ill-defined, we must consider  $n$ odd.      

    Using the results     in Propositions \ref{pitchfork}  and \ref{unstable_criterion}, we  deduce:
 
\begin{proposition}
Consider the  homographic dynamics of the 3-d gravitational $n$-body problem on   $\mathbb{S}^2$, where $n$ is odd. 
If
\begin{equation}
2\pi   \sum \limits_{k=1}^{(n-1)/2}  \left(1+ \cot^2\left(\frac{2k\pi}{n}  \right)   \right) 
\left(
3n - 8k(n-k)  \pi\cot\left(\frac{2k\pi}{n}  \right) \right) > 0\,
\label{sphere_again}
\end{equation}
 then the Equatorial Lagrangian homographic RE   experiences a subcritical pitchfork bifurcation.
Moreover, denoting the bifurcation momentum absolute value  $c_0$,  the Equatorial Lagrangian homographic RE are unstable  for  $|c|<c_0$.

\end{proposition}

\noindent
Proof.   The sphere $\mathbb{S}^2$ is described by a parametrization  \eqref{param_M} with a generatrix $f(z)= \sqrt{1-z^2}$ and the Equator corresponds to the circle $z=0$. The unit sphere  $\mathbb{S}^2$  has constant positive curvature $K=1$; in particular, in our  parametrization,    $f''(z)<0$ for all $z.$  
 Since  $G(x) = -\cot(x)$ we have $G'(x) = 1+ \cot^2 x>0$  and $G''(x) =  -2(1+ \cot^2 x) \cot x$. 
%
%
By Proposition \ref{pitchfork} and Remark \ref{inner_branch}, the presence of a subcritical pitchfork is insured by the condition \eqref{subcritical_pitch} written as a strict inequality, that is
\begin{equation}
\sum \limits_{k=1}^{n-1} 
2(n-k)k\pi \left( 1+ \cot^2   \left(\frac{2k\pi}{n}  \right) \right)
  \left(3 - 4k\pi\cot\left(\frac{2k\pi}{n}\right) \right) >  0\,.
  \label{sphere_condition}
\end{equation}
Recall that $n$ must be odd. Expanding the sum above and since $\cot(2\pi-x) = \cot x$ we have
\begin{align}
&\sum \limits_{k=1}^{n-1} 
2(n-k)k\pi \left( 1+ \cot^2   \left(\frac{2k\pi}{n}  \right) \right)
  \left(3 - 4k\pi\cot\left(\frac{2k\pi}{n}\right) \right)  \nonumber  \\
  &= 
  \sum \limits_{k=1}^{(n-1)/2} 2\pi \left(1+ \cot^2\left(\frac{2k\pi}{n}  \right)   \right)
  \left(
  3(n-k) - 4(n-k)  k\pi \cot \left(\frac{2k\pi}{n}  \right) + 3k - 4k(n-k)  \pi \cot \left(\frac{2(n-k)\pi}{n}  \right)
  \right)  \nonumber \\
  & =
2\pi   \sum \limits_{k=1}^{(n-1)/2}  \left(1+ \cot^2\left(\frac{2k\pi}{n}  \right)   \right) 
\left(
3n - 8k(n-k)  \pi\cot\left(\frac{2k\pi}{n}  \right)
\right)
\end{align}
and thus condition \eqref{sphere_again} is obtained.  
The lack of stability of the inner branch of the  RE follows as  an application of from Proposition \ref{unstable_criterion}.
$\square$
 
  \begin{remark}
The bifurcation value is the positive root of 
 \begin{equation}
c_0^2 = \frac{1}{n}\, \sum \limits_{k=1}^{n-1} \frac{2(n-k)k\pi}{n}\left( 1+ \cot^2   \left(\frac{2k\pi}{n}  \right) \right)\,.
\label{c_0-sphere}
\end{equation}
  \end{remark}
 
 \begin{remark}
 One may verify that condition \eqref{sphere_condition} is  satisfied for $n=3$. Moreover, one may verify that the sum in \eqref{sphere_condition} is strictly positive and so by Remark \ref{inner_branch} we have a subcritical pitchfork.
  \end{remark}
 
  \begin{conjecture}
  \label{conjecture}
In  the 3-d gravitational $n$-body problem on   $\mathbb{S}^2$ the Equatorial Lagrangian homographic RE   experiences a subcritical pitchfork bifurcation. %
 %
 \end{conjecture}
 
 \begin{remark}
 To prove the conjecture above it is sufficient to prove the inequality \eqref{sphere_again} for any odd $n$.
 \end{remark}

Likewise, we have

\begin{proposition}
\label{H_2_proposition}
Consider the  homographic dynamics of the 3-d gravitational $n$-body problem on   $\mathbb{H}^2_{\text{one}}$.  
If 
%
%
\begin{equation}
 \sum \limits_{k=1}^{n-1} 
2(n-k)k\pi\left( \coth^2 \left( \frac{2k\pi}{n} \right)  -1 \right)  \left[
3 +4k\pi
\coth \left( \frac{2k\pi}{n} \right)
\right]
 > 0
  \label{hyp_condition}
\end{equation}
 then the Equatorial Lagrangian homographic RE   experiences a subcritical pitchfork bifurcation. Moreover, denoting the bifurcation momentum absolute value  $c_0$, then Equatorial Lagrangian homographic RE is unstable for  $|c|>c_0$.

\end{proposition}

\begin{figure}[h!]
\centering
      \subfigure [The amended potential $W_c(z)$ for $n=3$ for the  3-d gravitational potential $G(x)=-\cot x$ on  $\mathbb{S}^2$.]
       {
            \includegraphics[angle=0,scale=0.26] {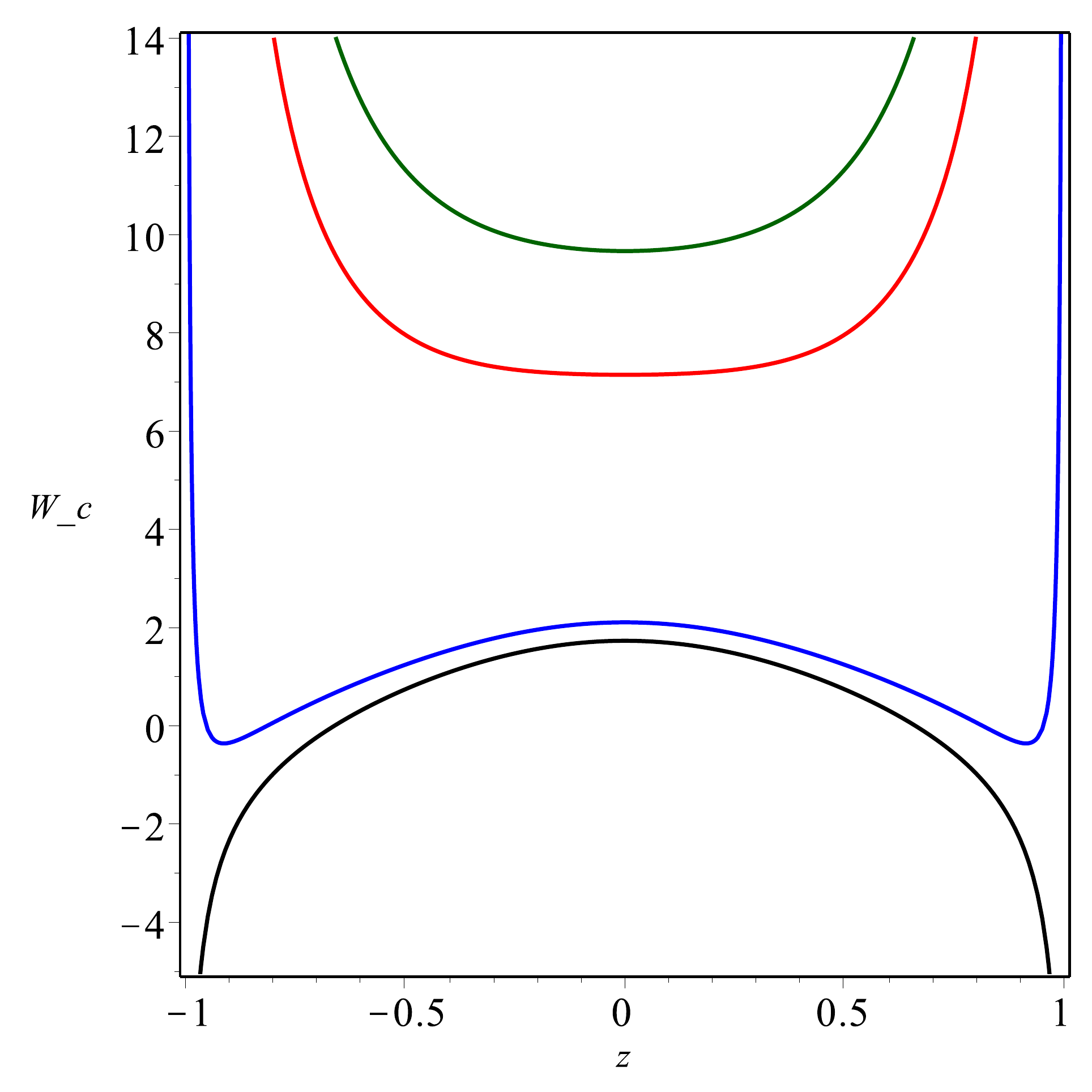}
         } \quad \quad \quad 
         \subfigure [The amended potential $W_c(z)$ for $n=3$ for 3-d gravitational potential $G(x)=-\coth x $ on $\mathbb{H}^2_{\text{one}}$.]
       {
            \includegraphics[angle=0,scale=0.257] {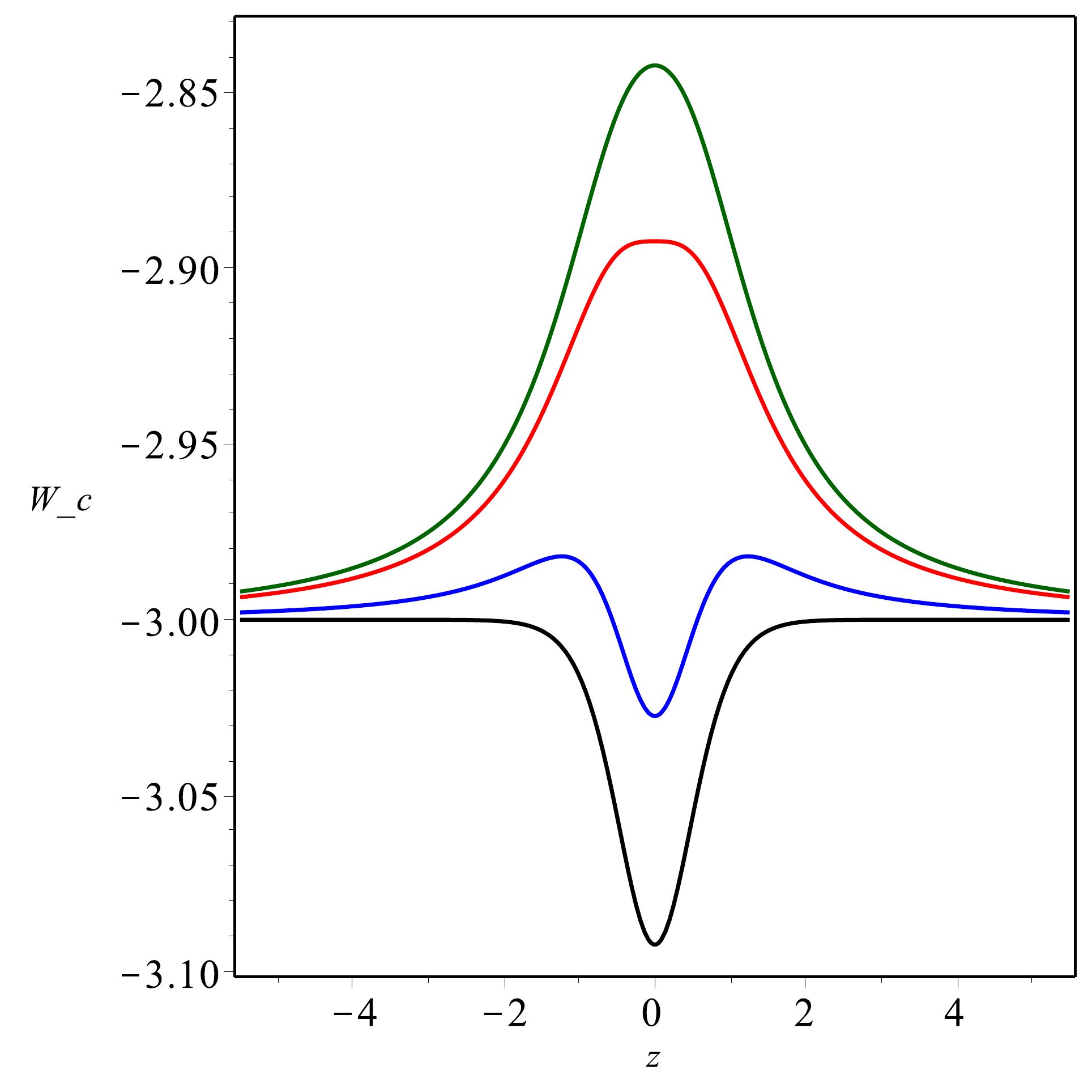}
         } 
         \caption{The amended potential $W_c(z)$  for motion  on $\mathbb{S}^2$ and  $\mathbb{H}^2_{\text{one}}$ with the  gravitational potential for $n=3$. 
         The black, blue, red and green curves correspond to momenta  $c=0$, $|c|\in (0, c_0)$, $|c|=c_0$ and $|c|>c_0$, respectively. }
      \label{W_c_coth}
\end{figure}

  The sketch the amended potential for homographic motion on $\mathbb{S}^2$ and $\mathbb{H}^2_{\text{one}}$  for $n=3$ is given in Figure  \ref{W_c_coth}.
 %
   The orbits on  $\mathbb{S}^2$ are qualitatively identical to the the quasi-harmonic homographic orbits, with one modification: 
since the potential is not defined at the Poles,  orbits reaching these points correspond to an ejection/fall from/into a collision configuration.  The dynamics are described in detail in \cite{Diacu2011}. 
%
%

%
%
%
%



%
%

 \begin{figure}[h!]
\centering
      \subfigure[The phase curves for  $c=0.$]
      {
           \includegraphics[angle=0,scale=0.25] {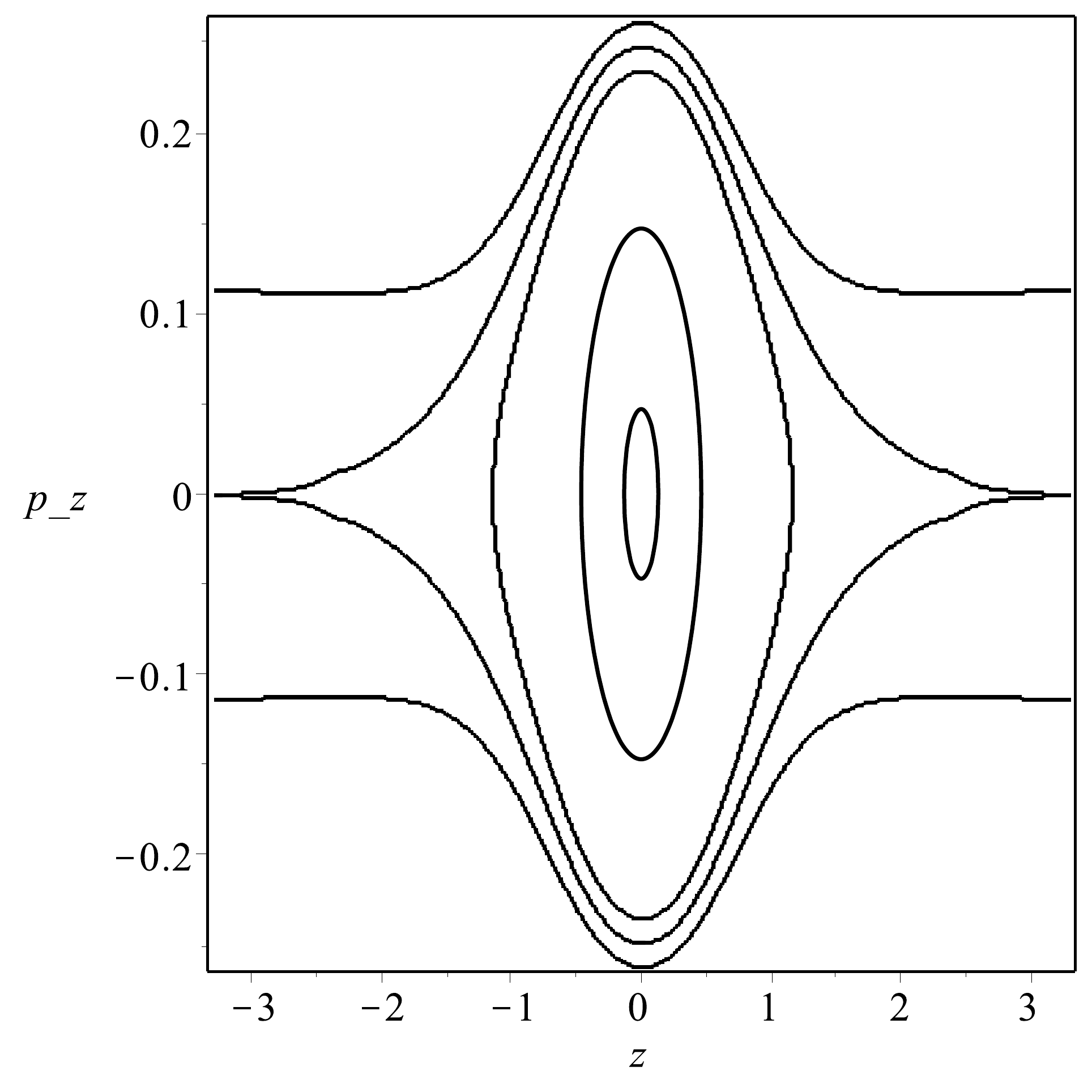}
           } \quad 
      \subfigure [The phase curves for  $|c| \in (0, c_0).$]
       {
            \includegraphics[angle=0,scale=0.25] {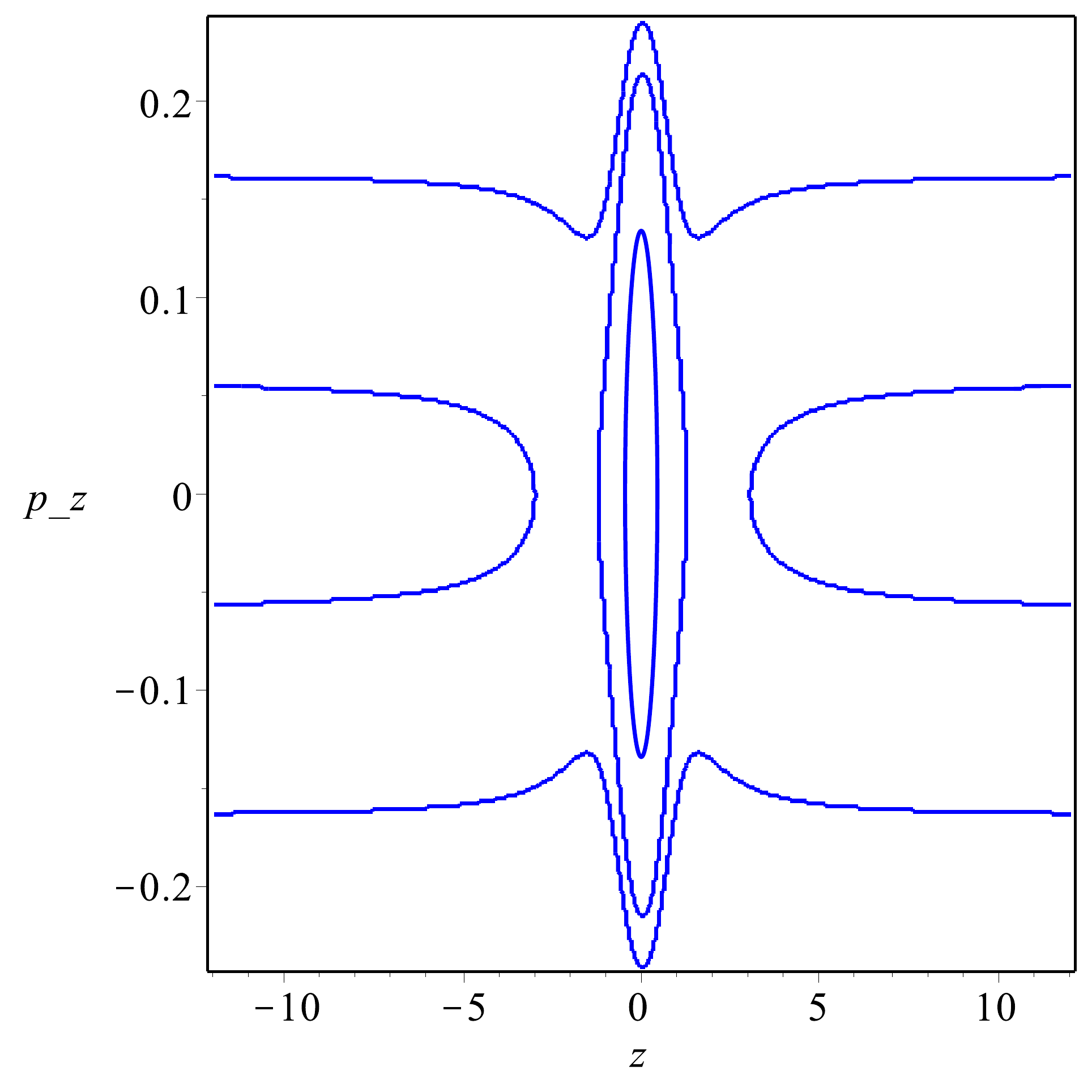}
         } \quad 
         \subfigure [The phase curves for  $|c| \geq c_0.$]
       {
            \includegraphics[angle=0,scale=0.25] {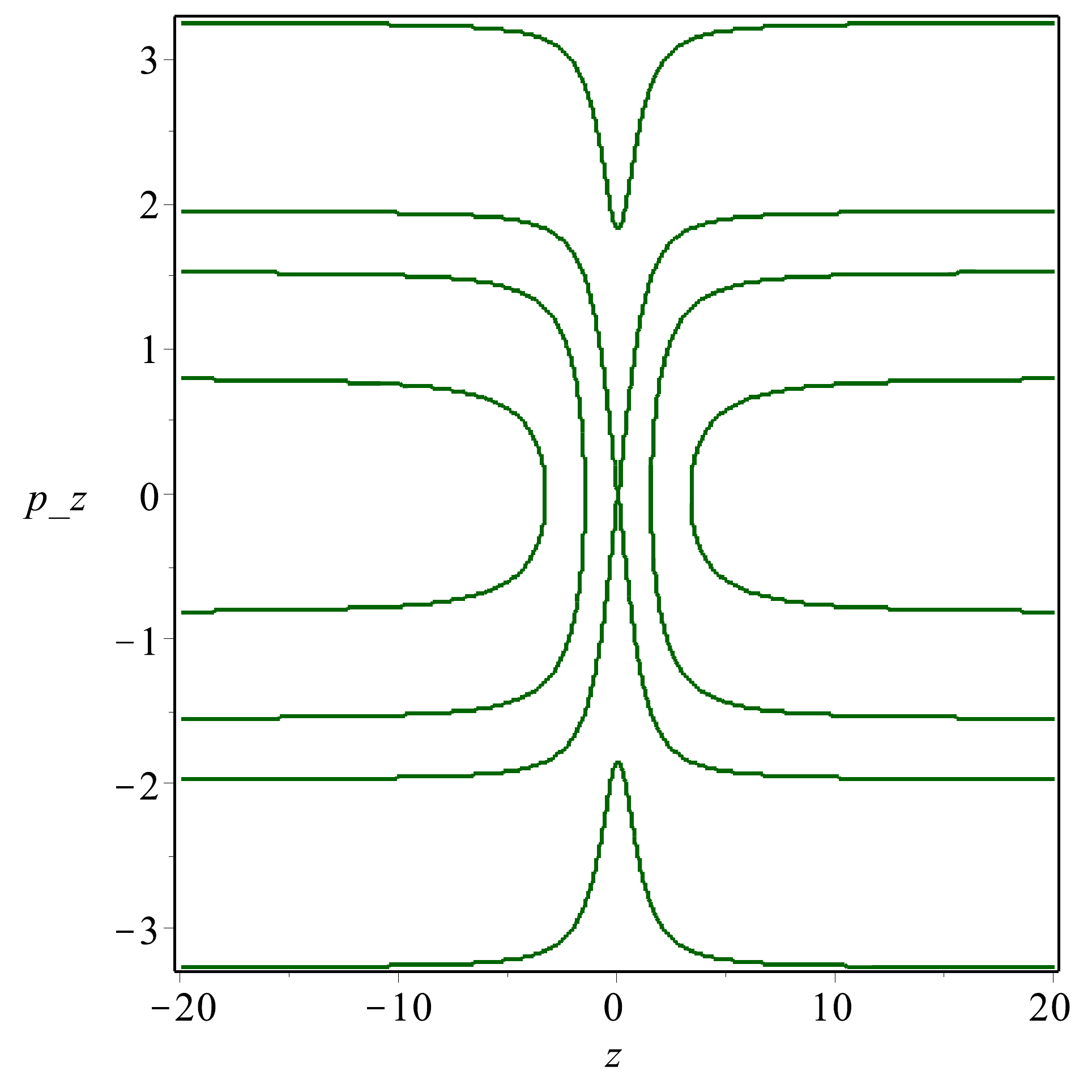}
         } 
         \caption{The  phase plane of the reduced dynamics  on the homographic invariant manifold for motion on $\mathbb{H}^2_{\text{one}}$ with the  gravitational potential for $n=3$.  
}
         \label{Figure_flow_hyper}
\end{figure}
 %
 %
 We now comment on the  dynamics  for the hyperbolic case.   The amended potential is
\begin{align}
 W_c(z) = \frac{3c^2}{2(1+z^2)} -  3 \coth\left( \frac{2\pi}{3} \sqrt{1+z^2}\right)\,,
 \label{amended _hyper}
\end{align}
and further we calculate
\begin{align}
 W'_c(z) = z\left(3c^2 - 2\pi \left( 1-\coth^2 \left( \frac{2\pi}{3}\right) \right) \right)\,.
 \end{align}
There is one critical point at $z=0$, and  
the bifurcation momentum is
\begin{align}
c_0:= \sqrt{\frac{2\pi}{3} \left( \coth^2\left(  \frac{2\pi}{3} -1\right) \right)}\simeq 0.1309853861\,.
\label{crit_c_hyper}
\end{align}
%
%
%
%
Applying Proposition \ref{H_2_proposition}, or just by direct inspection of the profile of $W_c(z)$, we deduce the following:

\begin{itemize}

\item for $|c|=0$ there is one equilibrium and it is located on the Equator;

\item for $|c|<c_0$ there are three RE, one with its trajectory on the Equator and two with their trajectories on the circles $z=\pm z (c)$, the later satisfying
\begin{equation}
\frac{3c^2}{(1+z_c^2)} -  \frac{2\pi \left( 1-\coth^2 \left( \frac{2\pi}{3} \sqrt{1+z (c)^2}\right) \right)}{\sqrt{1+z (c)^2}} =0\,;
\end{equation}

\item  at $|c|=c_0$  there is a pitchfork bifurcation;

\item for $|c|>c_0$ there is a unique  RE with its trajectory on the Equator.

\end{itemize}

The orbit behaviour and the topology of the phase space may be deduced following an analogous procedure as in the previous subsection. We sketch the phase portrait for representative momentum levels in Figure  \ref{Figure_flow_hyper}. 




\section{Acknowledgements} The author was supported by  NSERC Discovery grant. She thanks Florin Diacu  and Stefanella Boatto for suggesting this problem. She also thanks Tanya Schmah  for relevant discussions.


\end{document}